\documentclass[aps,prl,preprintnumbers,showpacs,nofootinbib,floatfix]{revtex4}
\usepackage[english]{babel}
\usepackage{graphicx}
\usepackage{subfig}
\usepackage{dcolumn}
\usepackage{bm}
\usepackage{enumerate}
\usepackage{subfig}
\usepackage{rotating}
\usepackage{multirow}
\usepackage{psfrag}
\usepackage{gensymb}
\usepackage{amsmath}
\usepackage{amsfonts}
\usepackage{amssymb}
\usepackage{latexsym}
\begin{document}
\newcommand{\newc}{\newcommand}
\newc{\ra}{\rightarrow}
\newc{\lra}{\leftrightarrow}
\newc{\beq}{\begin{equation}}
\newc{\eeq}{\end{equation}}
\newc{\barr}{\begin{eqnarray}}
\newc{\earr}{\end{eqnarray}}
\def \lta {\mathrel{\vcenter
     {\hbox{$<$}\nointerlineskip\hbox{$\sim$}}}}
\def \gta {\mathrel{\vcenter
     {\hbox{$>$}\nointerlineskip\hbox{$\sim$}}}}
\def\vbf{\mbox{\boldmath $\upsilon$}}
\def\barr{\begin{eqnarray}}
\def\earr{\end{eqnarray}}
\def\g{\gamma}
\newcommand{\dphi}{\delta \phi}
\newcommand{\bupsilon}{\mbox{\boldmath \upsilon}}
\newcommand{\at}{\tilde{\alpha}}
\newcommand{\pt}{\tilde{p}}
\newcommand{\Ut}{\tilde{U}}
\newcommand{\rhb}{\bar{\rho}}
\newcommand{\pb}{\bar{p}}
\newcommand{\pbb}{\bar{\rm p}}
\newcommand{\kt}{\tilde{k}}
\newcommand{\kb}{\bar{k}}
\newcommand{\wt}{\tilde{w}}
\title{Predicted modulated  differential rates for direct WIMP searches at low energy transfers}
\author{ J. D. Vergados \footnote{vergados@uoi.gr}}
\affiliation{CERN, Theory Division, CH 1211, Geneva 23, Switzerland\\ and\\ Theoretical Physics Division, University
of Ioannina, Ioannina, Gr 451 10, Greece}
 \begin{abstract}
The differential event rate for direct detection of dark matter, both the time averaged and the modulated one due to the motion of the Earth, are discussed. The calculations focus on relatively light cold dark matter candidates  (WIMP) and low energy transfers. It is shown that for sufficiently light WIMPs the extraction of relatively large nucleon cross sections is possible. Furthermore for some WIMP masses the modulation amplitude may change sign, meaning that, in such a case, the maximum rate may occur six months later than naively expected. This effect can be exploited to yield information about the mass of the dark matter candidate, if and when the observation of the modulation of the event rate is established.
 \end{abstract}

\date{\today}
\maketitle
keywords:{
Dark matter, WIMP,  direct  detection,  WIMP-nucleus scattering, event rates, modulation}\\

\section{Introduction}
The combined MAXIMA-1 \cite{MAXIMA-1}, BOOMERANG \cite{BOOMERANG},
DASI \cite{DASI} and COBE/DMR Cosmic Microwave Background (CMB)
observations \cite{COBE} imply that the Universe is flat
\cite{flat01}
and that most of the matter in
the Universe is Dark \cite{SPERGEL},  i.e. exotic. These results have been confirmed and improved
by the recent WMAP data \cite{WMAP06}. Combining the
the data of these quite precise experiments one finds:
$$\Omega_b=0.0456 \pm 0.0015, \quad \Omega _{\mbox{{\tiny CDM}}}=0.228 \pm 0.013 , \quad \Omega_{\Lambda}= 0.726 \pm 0.015.$$
Since any "invisible" non exotic component cannot possibly exceed $40\%$ of the above $ \Omega _{\mbox{{\tiny CDM}}}$
~\cite {Benne}, exotic (non baryonic) matter is required and there is room for cold dark matter candidates or WIMPs (Weakly Interacting Massive Particles).

Even though there exists firm indirect evidence for a halo of dark matter
in galaxies from the
observed rotational curves, see e.g the review \cite{UK01}, it is essential to directly
detect
such matter.
Until dark matter is actually detected, we shall not be able to
exclude the possibility that the rotation curves result from a
modification of the laws of nature as we currently view them.  This makes it imperative that we
invest a
maximum effort in attempting to directly detect dark matter in the laboratory. Furthermore such a direct detection will also
unravel the nature of the constituents of dark matter.
The possibility of such detection, however, depends on the nature of the dark matter
constituents and their interactions.

Since the WIMP's are  expected to be
extremely non relativistic, with average kinetic energy $\langle T\rangle  \approx
50 \ {\rm keV} (m_{\mbox{{\tiny WIMP}}}/ 100 \ {\rm GeV} )$, they are not likely to excite the nucleus, even if they are quite massive $m_{\mbox{{\tiny WIMP}}} > 100$ GeV.
So they can be directly detected mainly via the recoiling of a nucleus
(A,Z) in elastic scattering. The event rate for such a process can
be computed from the following ingredients:
i) An effective Lagrangian at the elementary particle (quark)
level obtained in the framework of the prevailing particle theory.  In supersymmetry  the dark matter candidate is the
LSP (Lightest Supersymmetric Particle) \cite{ref2a,ref2b,ref2c,ref2,ELLROSZ,Gomez,ELLFOR}.
In this case the effective Lagrangian is constructed as described,
e.g., in Refs.~\cite{ref2a,ref2b,ref2c,ref2,ELLROSZ,Gomez,ELLFOR,JDV96,JDV06a}. ii)
A well defined procedure for transforming the amplitude thus
obtained, using the previous effective Lagrangian, from the quark to
the nucleon level. To achieve this one needs a quark model for the nucleon, see e.g.  \cite{JDV06a,Dree00,Dree,Chen}. This step is particularly important in supersymmetry or other models dominated by a scalar interaction (intermediate Higgs etc), since, then, the elementary amplitude becomes proportional to the quark mass and the content of the nucleon in quarks other than $u$ and $d$  becomes very important.
iii) knowledge of the relevant nuclear matrix elements
\cite{Ress,DIVA00}, obtained with as reliable as possible many
body nuclear wave functions, iv) knowledge of the WIMP density in our vicinity and its velocity distribution.

From steps i) and ii) one obtains the nucleon cross sections. These can also be extracted from the data of event rates, if and when such data become available. From limits on the event rates, one can obtain exclusion plots on the nucleon cross sections as  functions of the WIMP mass. The extracted cross sections depend, of course, on inputs from steps iii)-iv).

In the standard nuclear recoil experiments, first proposed more than 30 years ago \cite{GOODWIT}, one has to face the problem that the reaction of interest does not have a characteristic feature to distinguish it
from the background. So for the expected low counting rates the background is
a formidable problem. Some special features of the WIMP-nuclear interaction can be exploited to reduce the background problems. Such are:
i) the modulation effect: this yields a periodic signal due to the motion of the earth around the sun. Unfortunately this effect, also proposed a long time ago \cite{Druck} and subsequently studied by many authors \cite{PSS88,GS93,RBERNABEI95,LS96,ABRIOLA98,HASENBALG98,GREEN04,SFG06}, is small and becomes even smaller than  $2\%$ due to cancelations arising from nuclear physics effects,
ii) backward-forward asymmetry expected in directional experiments, i.e. experiments in which the direction of the recoiling nucleus is also observed. Such an asymmetry has also been predicted a long time ago \cite{SPERGEL88}, but it has not been exploited, since such experiments have been considered  very difficult to perform, but they now appear to be feasible \cite{SPERGEL88,DRIFT,SHIMIZU03,KUDRY04,DRIFT2,GREEN05,Green06,KRAUSS,KRAUSS01,Alenazi08,Creswick010,Lisanti09}.
iii) transitions to excited states: in this case one need not measure nuclear recoils, but the de-excitation $\gamma$ rays. This can happen only in very special cases since the average WIMP energy is too low to excite the nucleus. It has, however, been found that in the special case of the target $^{127}$I such a process is feasible \cite{VQS04} with branching ratios around $5\%$,
(iv) detection of electrons produced during the WIMP-nucleus collision:
it turns out, however, that this production peaks at very low energies. So only gaseous TPC detectors can reach the desired level of $100$ eV. In such a case the number of electrons detected may exceed the number of recoils for a target with high $Z$ \cite{VE05,MVE05},
v) detection of hard X-rays produced when the inner shell holes are filled:
it has been found \cite{MouVerE} that in the previous mechanism inner shell electrons can be ejected. These holes can be filled by the Auger process or X-ray emission.

 In connection with nuclear structure aspects, in a series of calculations, e.g. in \cite{JDV03,JDVSPIN04,VF07} and references there in, it has been shown that for the coherent contribution, due to the scalar interaction, the inclusion of the nuclear form factor is important, especially in the case of relatively heavy targets. They also showed that the nuclear spin cross sections  are characterized by a single, i.e. essentially isospin independent, structure function and two static spin values, one for the  proton and one for the neutron, which depend on the target.
 
  As we have already mentioned an essential ingredient in direct WIMP detection is the WIMP density in our vicinity and, especially, the WIMP velocity distribution. Some of the calculations have considered various forms of phenomenological non symmetric velocity distributions  \cite{DRIFT2,GREEN04,GREEN05,SFG06} and some of them even more exotic dark matter flows like
the late infall of dark matter into  the galaxy, i.e caustic rings
 \cite{SIKIVI1,SIKIVI2,Verg01,Green,Gelmini}, dark matter orbiting the
 Sun \cite{KRAUSS} and Sagittarius dark matter \cite{GREEN02}.  
 
  In addition to computing the time averaged rates, these calculations studied the modulation effect. They showed that in the standard recoil experiments the modulation amplitude  in the total rate may change sign for large reduced mass, i.e. heavy WIMPS and large A.
In directional experiments, in addition to the expected asymmetry mentioned above, the modulation exhibits two very interesting patterns i) its magnitude in certain directions can be very large and ii) the location of the maximum and minimum depends on the direction of observation.

In the present paper we will expand the above calculations and study the differential event rates, both time averaged and modulated,  in the region of low energy transfers, as in the DAMA experiment \cite{DAMA1,DAMA11}, focusing our attention on relatively light WIMPS \cite{XENON10,CoGeNT11,FPSTV11}. Such light WIMPs  can be accommodated in some SUSY models \cite{CKWY11}. We will focus here on the standard Maxwell-Boltzmann (M-B)
distribution for the WIMPs of our galaxy and we will not be concerned with other  distributions
\cite{VEROW06,JDV09,TETRVER06,VSH08},
even though some of them
 may affect the modulation. The latter will be studied elsewhere. In such a context we will explicitly show that the modulation amplitude, entering both the differential and the total rates, changes sign for certain WIMP masses. As a result such an effect, if and when the needed data become available, may be exploited to infer the WIMP mass.


\section{The formalism for the WIMP-nucleus differential event rate}
This formalism adopted in this work is well known (see e.g. the recent reviews \cite{JDV06a,VerMou11}). So we will briefly discuss its essential elements here.
The differential event rate can be cast in the form:
\beq
\frac{d R}{ d Q}|_A=\frac{dR_0}{dQ}|_A+\frac{d{\tilde H}}{dQ}|_A \cos{\alpha}
\eeq
\barr
\frac{d R_0}{ d Q}|_A&=&\frac{\rho_{\chi}}{m_{\chi}}\frac{m_t}{A m_p} \sigma_n\left (\frac{\mu_r}{\mu_p} \right )^2 \sqrt{<\upsilon^2>} A^2\frac{1}{Q_0(A)}\frac{d t}{du}\nonumber\\
\frac{d {\tilde H}}{ d Q}|_A&=&\frac{\rho_{\chi}}{m_{\chi}}\frac{m_t}{A m_p} \sigma_n\left (\frac{\mu_r}{\mu_p} \right )^2 \sqrt{<\upsilon^2>} A^2 \frac{1}{Q_0(A)} \frac{d h}{du}
\label{drdu}
\earr
with with $\mu_r$ ($\mu_p$) the WIMP-nucleus (nucleon) reduced mass, $A$ is the nuclear mass number and $\sigma_n$ is the elementary WIMP-nucleon cross section. $ m_{\chi}$ is the WIMP mass and $m_t$ the mass of the target. The first term gives the time averaged rate, while the second gives the modulated amplitude. $\alpha$ is the phase of the earth ($\alpha=0$ on June 2nd). 

\begin{figure}
\begin{center}
\subfloat[]
{
\rotatebox{90}{\hspace{0.0cm} $\Psi_0(x)\rightarrow$}
\includegraphics[height=.17\textheight]{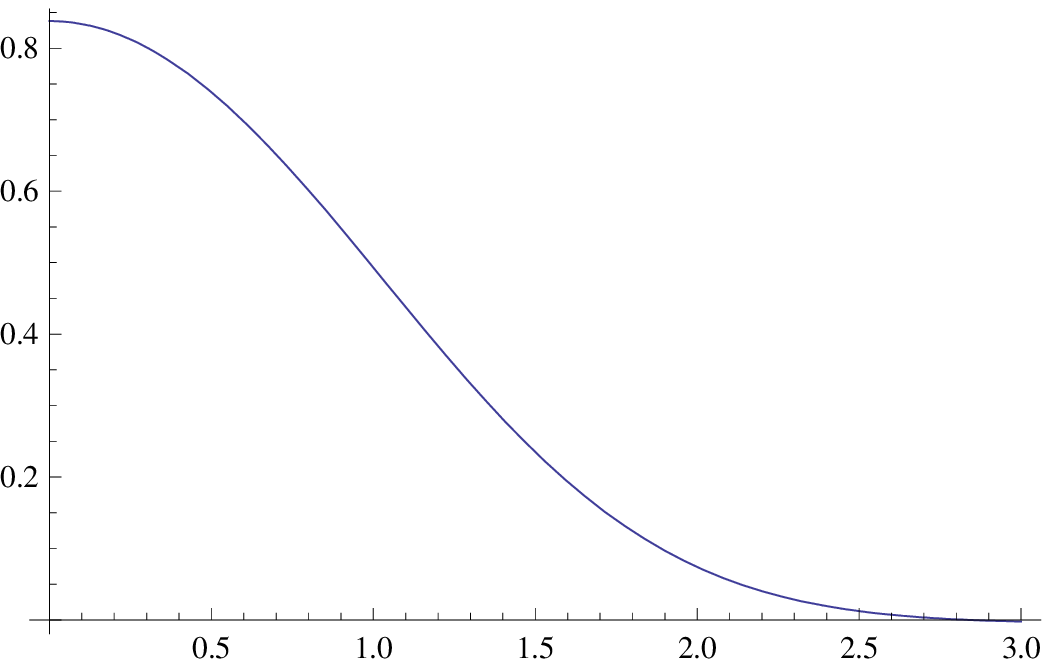}
}
\subfloat[]
{
\rotatebox{90}{\hspace{0.0cm} $\Psi_1(x)\rightarrow$}
\includegraphics[height=.17\textheight]{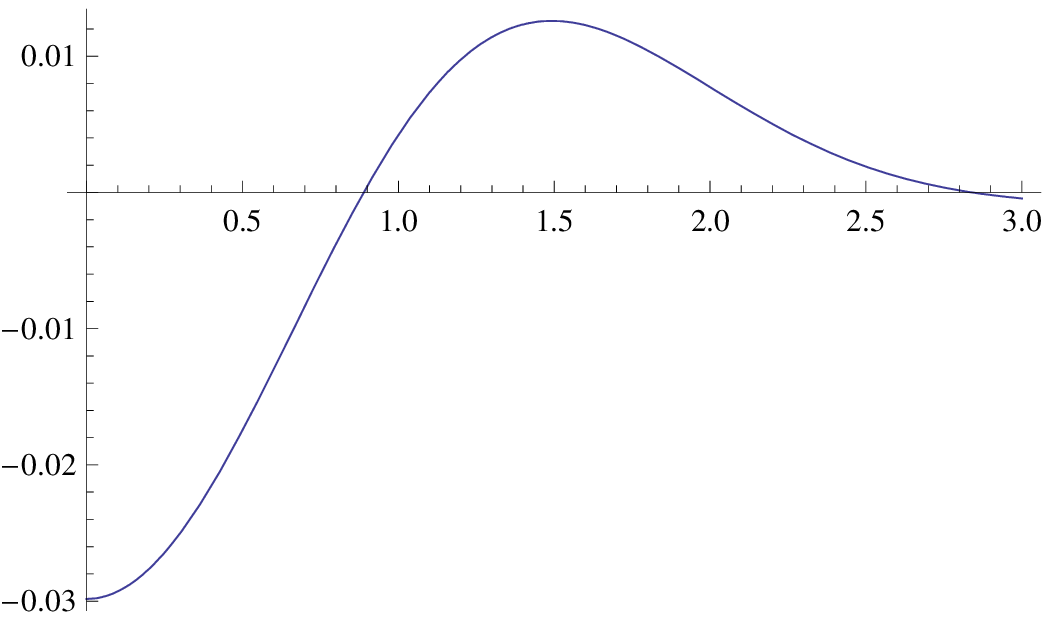}
}
\\
{\hspace{-2.0cm} $x=a \sqrt{u} \rightarrow$}
\caption{ The generic functions $\Psi_0(x)$  and $\Psi_1(x)$ entering   the differential rate, time averaged (a) and modulated (b). Note in (b) the change in sign at some point which depends on the target, the recoil energy  and the WIMP mass.
 \label{fig:psix}}
\end{center}
\end{figure}

\begin{figure}
\begin{center}
\subfloat[]
{
\rotatebox{90}{\hspace{0.0cm} $\Psi_0(a \sqrt{u})\rightarrow$}
\includegraphics[height=.17\textheight]{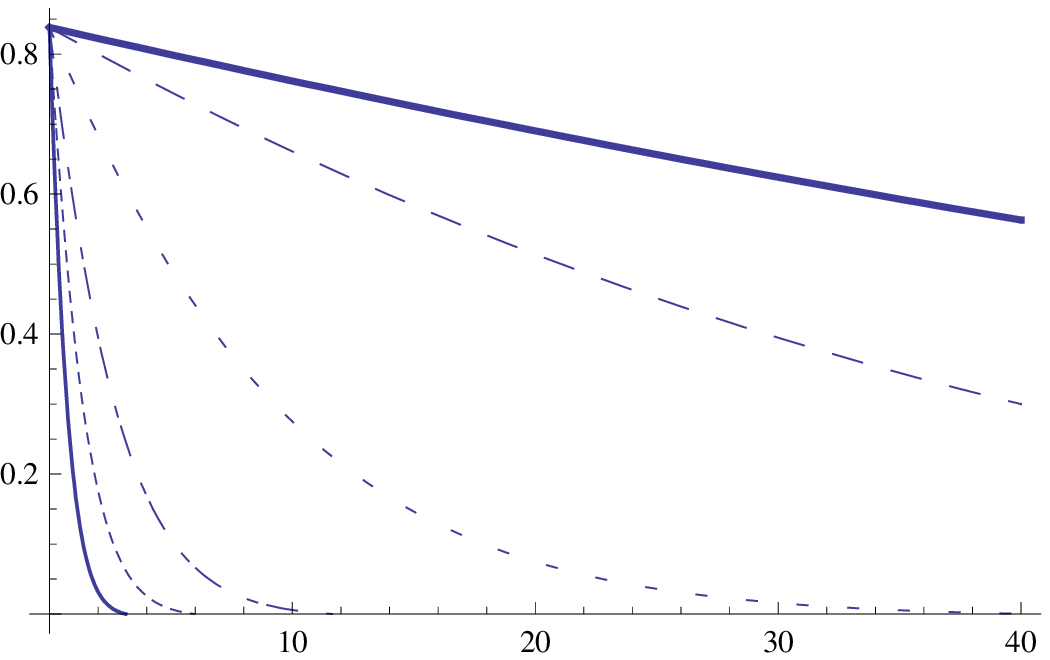}
}
\subfloat[]
{
\rotatebox{90}{\hspace{0.0cm} $\Psi_0(a \sqrt{u})F^2(u)\rightarrow$}
\includegraphics[height=.17\textheight]{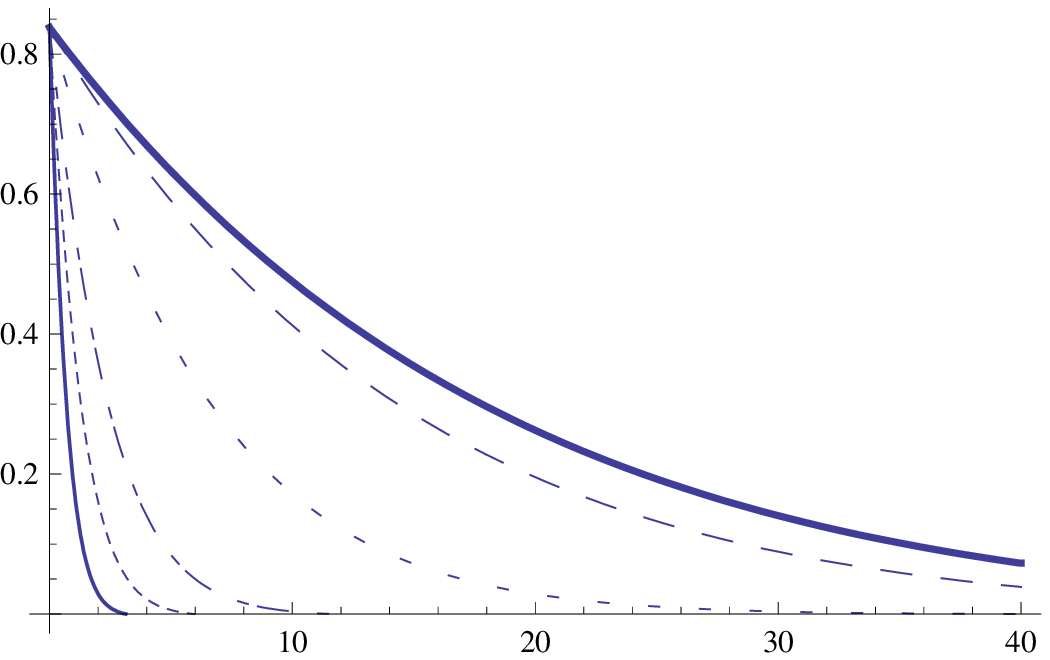}
}
\\
{\hspace{-2.0cm} $Q\rightarrow$keV}
\caption{ The function $\Psi_0(a \sqrt{u})$ entering   the differential rate as a function of the recoil energy for a heavy target, e.g. $^{127}$I, without the form factor (a) and including the form factor (b). The solid, dotted, dot-dashed, dashed, long dashed and thick solid lines correspond to 5, 7, 10, 20, 50 and 100 GeV WIMP masses.
 \label{fig:apsi0}}
\end{center}
\end{figure}

\begin{figure}
\begin{center}
\subfloat[]
{
\rotatebox{90}{\hspace{0.0cm} $\Psi_1(a \sqrt{u})\rightarrow$}
\includegraphics[height=.17\textheight]{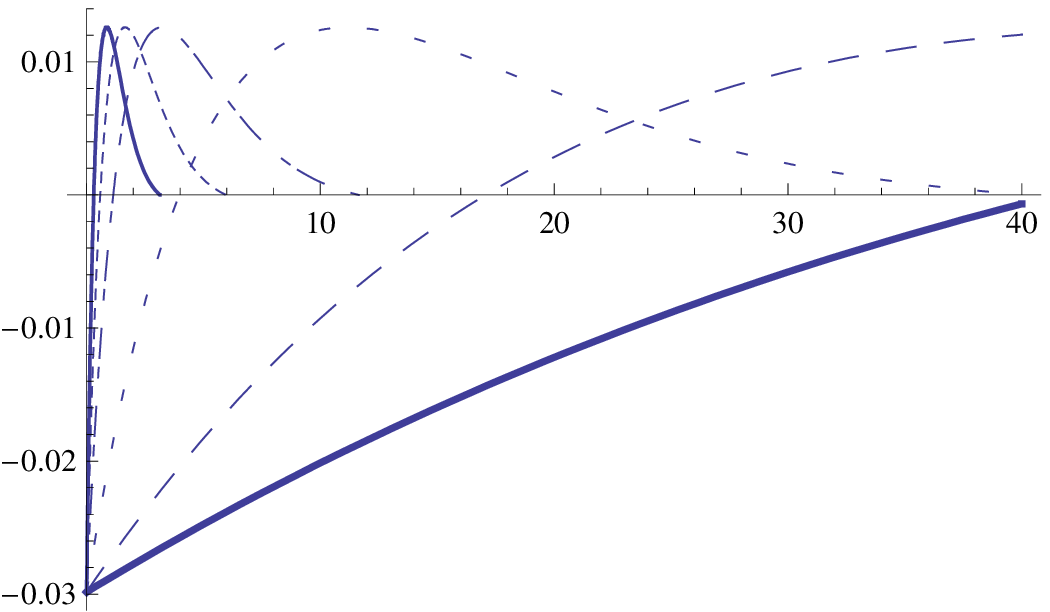}
}
\subfloat[]
{
\rotatebox{90}{\hspace{0.0cm} $\Psi_1(a \sqrt{u})F^2(u)\rightarrow$}
\includegraphics[height=.17\textheight]{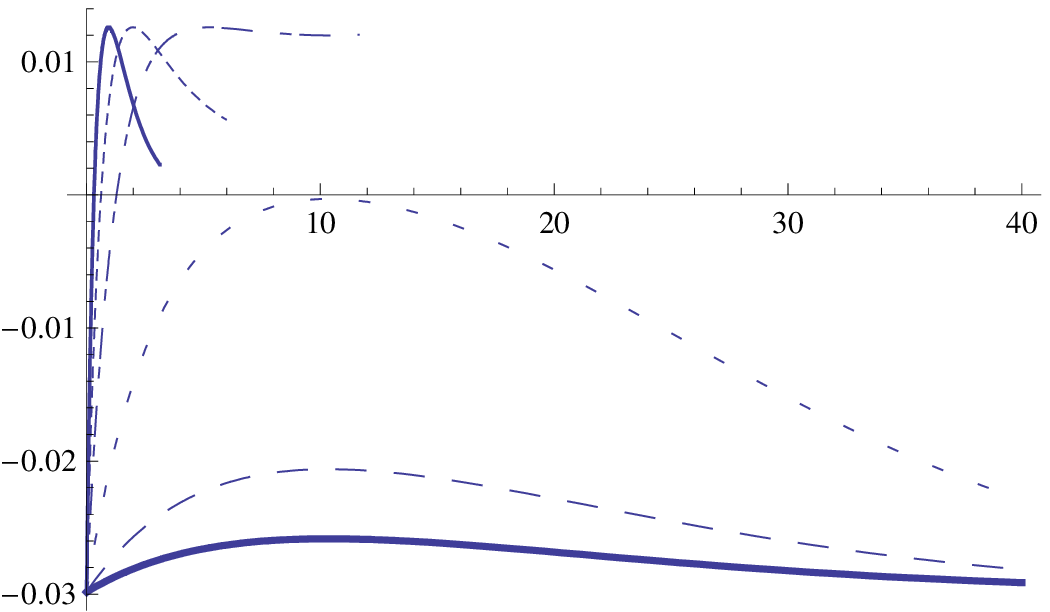}
}
\\
{\hspace{-2.0cm} $Q\rightarrow$keV}
\caption{ The function $\Psi_1(a \sqrt{u})$ entering   the modulated differential rate as a function of the recoil energy for a heavy target, e.g. $^{127}$I, without the form factor (a) and including the form factor (b). The solid, dotted, dot-dashed, dashed, long dashed and thick solid lines correspond to 5, 7, 10, 20, 50 and 100 GeV WIMP masses.
 \label{fig:apsi1}}
\end{center}
\end{figure}

\begin{figure}
\begin{center}
\rotatebox{90}{\hspace{0.0cm} $H(a \sqrt{u})\rightarrow$}
\includegraphics[height=.30\textheight]{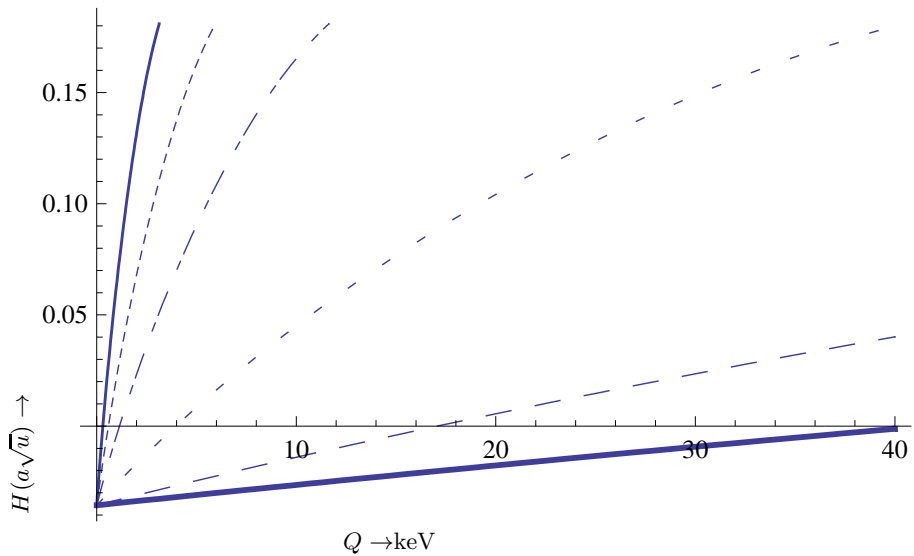}
\\
{\hspace{-2.0cm} $Q\rightarrow$keV}
\caption{ The same as in Fig. \ref{fig:apsi1} for function $H(a \sqrt{u})$ entering   the modulated differential rate as a function of the recoil energy for a heavy target, e.g. $^{127}$I. Note that this is independent of  the form factor. The solid, dotted, dot-dashed, dashed, long dashed and thick solid lines correspond to 5, 7, 10, 20, 50 and 100 GeV WIMP masses.
 \label{fig:apsiH}}
\end{center}
\end{figure}

Furthermore
\beq
\frac{d t}{d u}=\sqrt{\frac{2}{3}} a^2 F^2(u)   \Psi_0(a \sqrt{u}),\quad \frac{d h}{d u}=\sqrt{\frac{2}{3}} a^2 F^2(u) \Psi_1(a \sqrt{u})
\eeq
with $a=(\sqrt{2} \mu_r b \upsilon_0)^{-1}$, $\upsilon_0$ the velocity of the sun around the center of the galaxy and $b$ the nuclear harmonic oscillator size parameter characterizing the nuclear wave function.  $ u$ is the energy transfer $Q$ in dimensionless units given by
\begin{equation}
 u=\frac{Q}{Q_0(A)}~~,~~Q_{0}(A)=[m_pAb^2]^{-1}=40A^{-4/3}\mbox{ MeV}
\label{defineu}
\end{equation}
and $F(u)$ is the nuclear form factor. Note that the parameter $a$ depends both on the WIMP mass, the target and the velocity distribution. Note also that for a given energy transfer $Q$ the quantity $u$ depends on $A$.\\
The functions $\Psi_0(a \sqrt{u})$ and $\Psi_1(a \sqrt{u})$ can be obtained as follows:
\begin{itemize}
\item One starts with a Maxwell-Boltzmann distribution in the galactic frame with a characteristic velocity $\upsilon_0$ equal to the suns velocity around the center of the galaxy\footnote{Strictly speaking, since an upper cutoff is introduced to the velocity distribution, equal to the escape velocity, the velocity distribution should be renormalized. However the normalization integral is close to one, namely
\beq
\text{norm}=\frac{\sqrt{\pi } \text{erf}(y_{\text{esc}})-2
   e^{-y_{\text{esc}^2}} y_{\text{esc}}}{\sqrt{\pi }},\quad y_{\text{esc}}=\frac{\upsilon{y_{\text{esc}}}}{\upsilon_0},
\eeq
i.e. $\text{norm}\approx 0.9989$ for $y_{\text{esc}}=2.84$
}
\item one transforms to the local coordinate system:
\beq
{\bf y} \rightarrow {\bf y}+{\hat\upsilon}_s+\delta \left (\sin{\alpha}{\hat x}-\cos{\alpha}\cos{\gamma}{\hat y}+\cos{\alpha}\sin{\gamma} {\hat \upsilon}_s\right ) ,\quad y=\frac{\upsilon}{\upsilon_0}
\eeq
with $ {\hat \upsilon}_s$ a unit vector in the Sun's direction of motion and $\delta$ is the ratio of the Earth's velocity around the sun divided by $\upsilon_0$. The above formula assumes that the motion  of both the sun around the galaxy and of the Earth around the sun are uniformly circular. The exact orbits are, of course, more complicated \cite{GREEN04,LANG99}, but such deviations are not expected to significantly modify our results.
\item One integrates over the velocity integration over the angles and the result is multiplied the  velocity $y=\upsilon/\upsilon_0$ due to the WIMP flux.
\item The result is integrated from a minimum value, which depends on the energy transfer $y=a\sqrt{u}$, to a maximum $y=y_{\text{esc}}$, $y_{\text{esc}}=\upsilon_{\text{esc}}/ \upsilon_0$, $y_{\text{esc}}\approx2.84$
\end{itemize}
The result is
\barr
J(x)&=&\frac{1}{\delta  \cos \alpha-2}
\left [\text{erf}\left(-x+\frac{1}{2} \delta  \cos {\alpha
   }+1\right)+\text{erf}\left(x+\frac{1}{2} \delta  \cos
   {\alpha
   }+1\right)\right .\nonumber\\
 && \left .  +\text{erfc}\left(-y_{\text{esc}}+\frac{1}{2}
   \delta  \cos {\alpha
   }+1\right)
   +\text{erfc}\left(y_{\text{esc}}+\frac{1}{2}
   \delta  \cos {\alpha }+1\right)-2 \right ],\quad x=a\sqrt{u}
\earr
where erf$(x)$ and erfc$(x)$ are the error function and its complement respectively. Furthermore since $\delta=0.135$ we can expand in powers of $\delta$ and obtain:
\beq
J(a\sqrt{u})\approx\Psi_0(a \sqrt{u})+\Psi_1(a \sqrt{u})\cos{\alpha}+\Psi_2(a \sqrt{u})\cos{2 \alpha}
\eeq
with
\beq
\Psi_0(x)=\frac{1}{2}
   (\text{erf}(1-x)+\text{erf}(x+1)+\text{erfc}(1-y_{\text{esc}})+\text{erfc}(y_{\text{esc}}+1)-2)
\eeq
\barr
\Psi_1(x)&=&\frac{1}{4} \delta 
   \left( -\text{erf}(1-x)-\text{erf}(x+1)-\text{erfc}(1-y_{\text{esc}})-
   \text{erfc}(y_{\text{esc}}+1) \right . \nonumber\\
  && \left . +\frac{2 e^{-(x-1)^2}}{\sqrt{\pi }}
   +\frac{2
   e^{-(x+1)^2}}{\sqrt{\pi }}-\frac{2 e^{-(y_{\text{esc}}-1)^2}}{\sqrt{\pi
   }}-\frac{2 e^{-(y_{\text{esc}}+1)^2}}{\sqrt{\pi }}+2\right)
\earr
The function  $\Psi_2(x)$ is small, of order $\delta^2$ and it can be ignored. If, however, the experiments, which attempt to measure the modulation, want to go beyond the $\cos{\alpha}$ term, they should consider  terms $\cos{2\alpha}$ rather than $\sin{\alpha}$ as some of them have done.\\
 The functions $\Psi_0(x)$ and $\Psi_1(x)$ characterize both the coherent and the spin induced mode \cite{JDVSpin09}. We should note that the function $\Psi_1(x)$ changes sign at some value of $x$, which has implications on the total modulated rate, a point often missed (see Fig. \ref{fig:psix}).
Sometimes we will write the differential rate as:
\beq
\frac{d R}{ d Q}|_A=\frac{\rho_{\chi}}{m_{\chi}}\frac{m_t}{A m_p} \sigma_n \left ( \frac{\mu_r}{\mu_p} \right )^2 \sqrt{<\upsilon^2>} A^2 \frac{1}{Q_0(A)}\left(\frac{d t}{du}(1+ H(a \sqrt{u}) \cos{\alpha}\right )
\label{dhduH}
\eeq
In this formulation $H(a \sqrt{u}) $ gives the relative modulation amplitude (with respect to the time averaged one).
The functions $\Psi_0(a \sqrt{u})$ and $\Psi_1(a \sqrt{u})$, which exhibit the general characteristics of the differential rates, are exhibited in Figs \ref{fig:apsi0} and \ref{fig:apsi1}, while the function $H(a \sqrt{u})$ is shown in Fig. \ref{fig:apsiH}. These functions are independent of the nuclear physics. They only depend on the reduced mass and the velocity distribution. They are thus the same for both the coherent and the spin mode. Note that $\Psi_1(a \sqrt{u})$ and, consequently, $H(a \sqrt{u})$ can take both positive and negative values, which affects the location of the maximum.
\begin{figure}
\begin{center}
\includegraphics[height=.30\textheight]{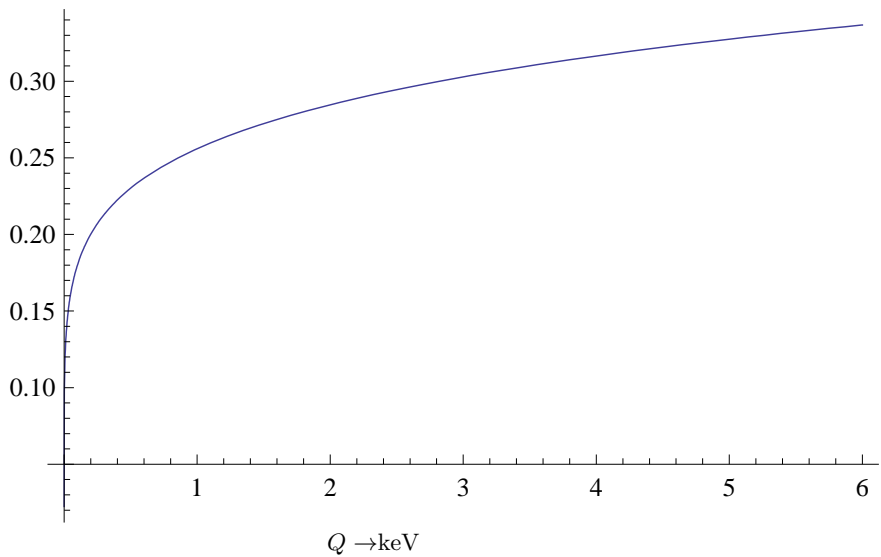}
\\
{\hspace{-2.0cm} $Q\rightarrow$keV}
\caption{ The quenching factor used in this work to transform keV$\rightarrow$keVee.
 \label{fig:quenchfac}}
\end{center}
\end{figure}
\begin{figure}
\begin{center}
\subfloat[]
{
\includegraphics[height=.25\textwidth]{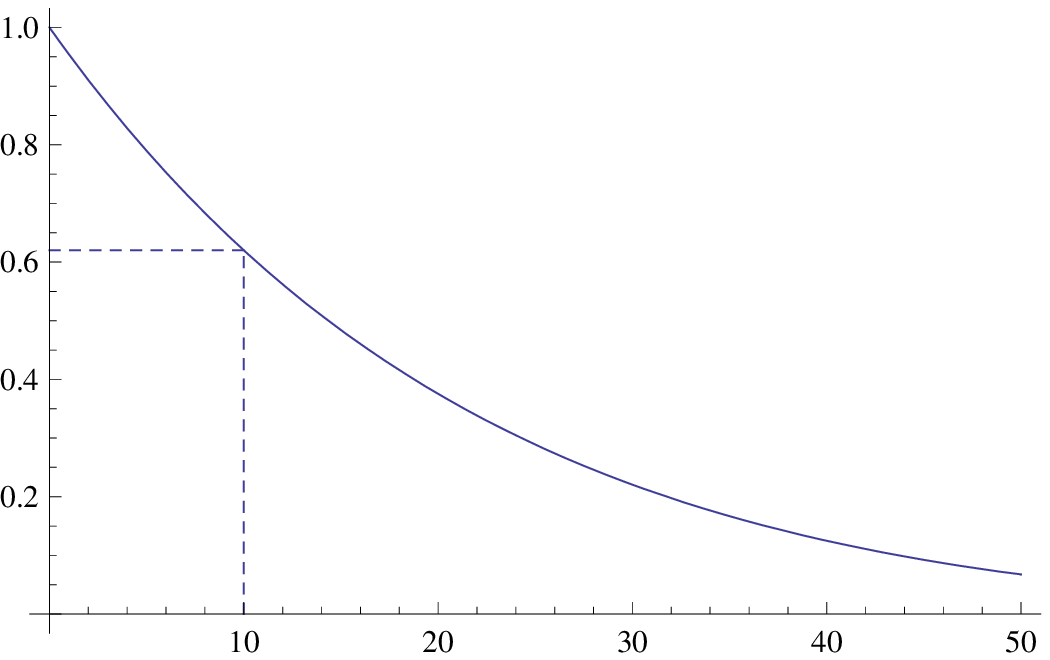}
}
\subfloat[]
{
\includegraphics[height=.25\textwidth]{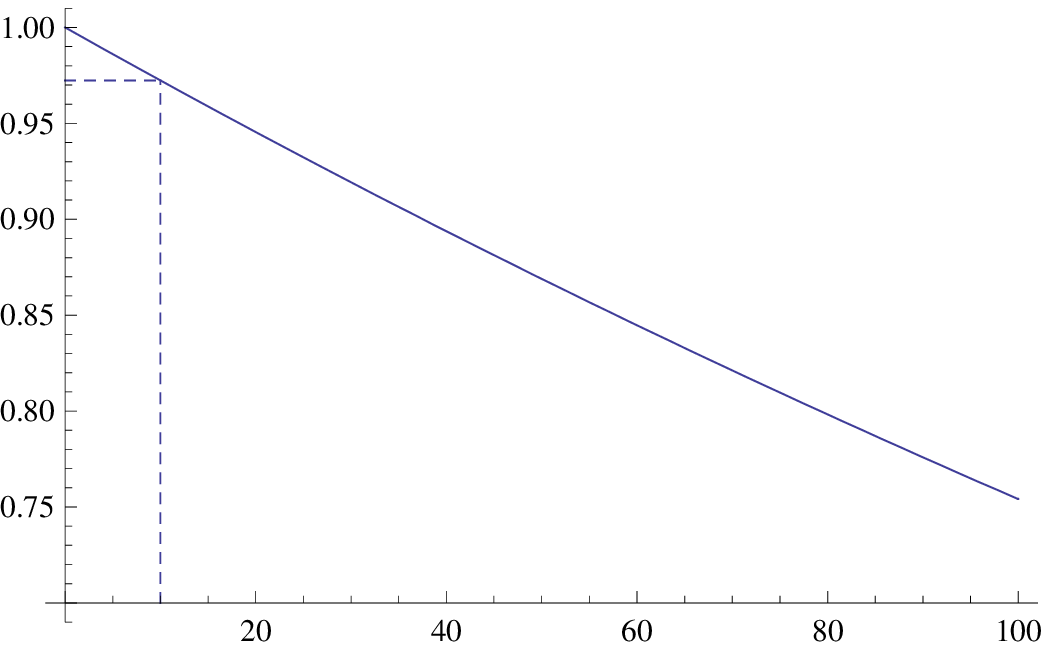}
}\\
{\hspace{-2.0cm} $Q\rightarrow$keV}
\caption{ The square of the nuclear form factor used in this work For $^{127}$I (a) and $^{23}$Na (b).
 \label{fig:formfac}}
\end{center}
\end{figure}
\section{Some results on differential rates}
\begin{figure}
\begin{center}
\subfloat[]
{
\rotatebox{90}{\hspace{0.0cm} $dR/dQ\rightarrow$kg/(y keVee)}
\includegraphics[height=.17\textheight]{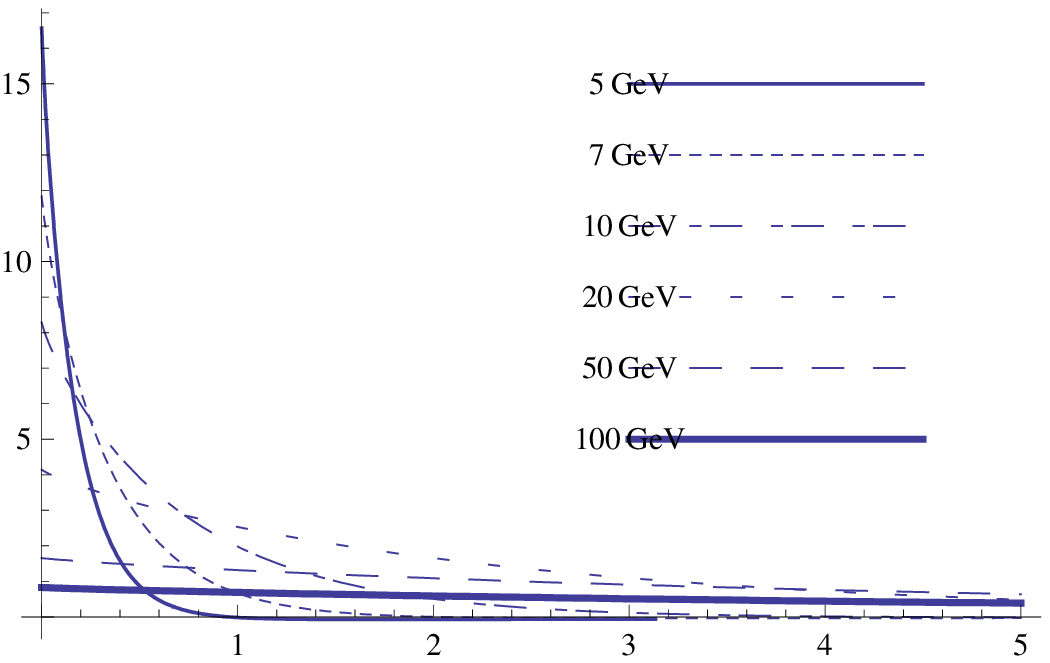}
}
\subfloat[]
{
\rotatebox{90}{\hspace{0.0cm} $d{\tilde H}/dQ\rightarrow$kg/(y keVee)}
\includegraphics[height=.17\textheight]{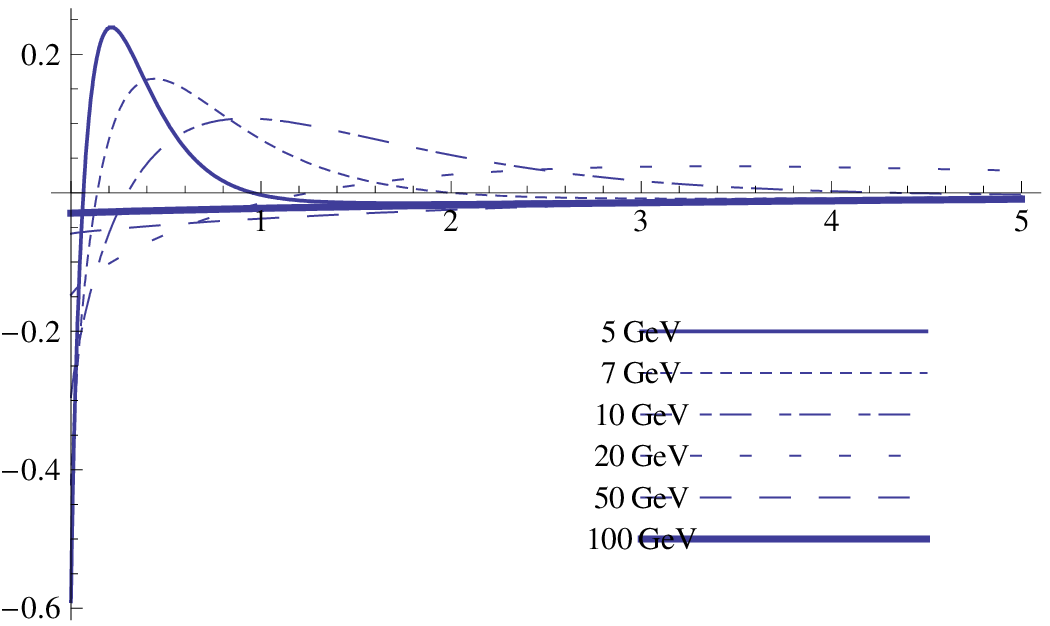}
}
\\
{\hspace{-2.0cm} $Q\rightarrow$keVee}
\caption{ The differential rate $\frac{dR}{dQ}$,   as a function of the recoil energy for a heavy target, e.g. $^{127}$I (a) and the amplitude for the modulated differential rate $\frac{d{\tilde H}}{dQ}$ (b), assuming a nucleon cross section of $10^{-7}$pb.  The solid, dotted, dot-dashed, dashed, long dashed and thick solid lines correspond to 5, 7, 10, 20, 50 and 100 GeV WIMP masses. Note that $\frac{d{\tilde H}}{dQ}$ is given in absolute units.
 \label{fig:dRdQdHdQ_127}}
\end{center}
\end{figure}

\begin{figure}
\begin{center}
\subfloat[]
{
\rotatebox{90}{\hspace{0.0cm} $dR/dQ\rightarrow$kg/(y keVee)}
\includegraphics[height=.17\textheight]{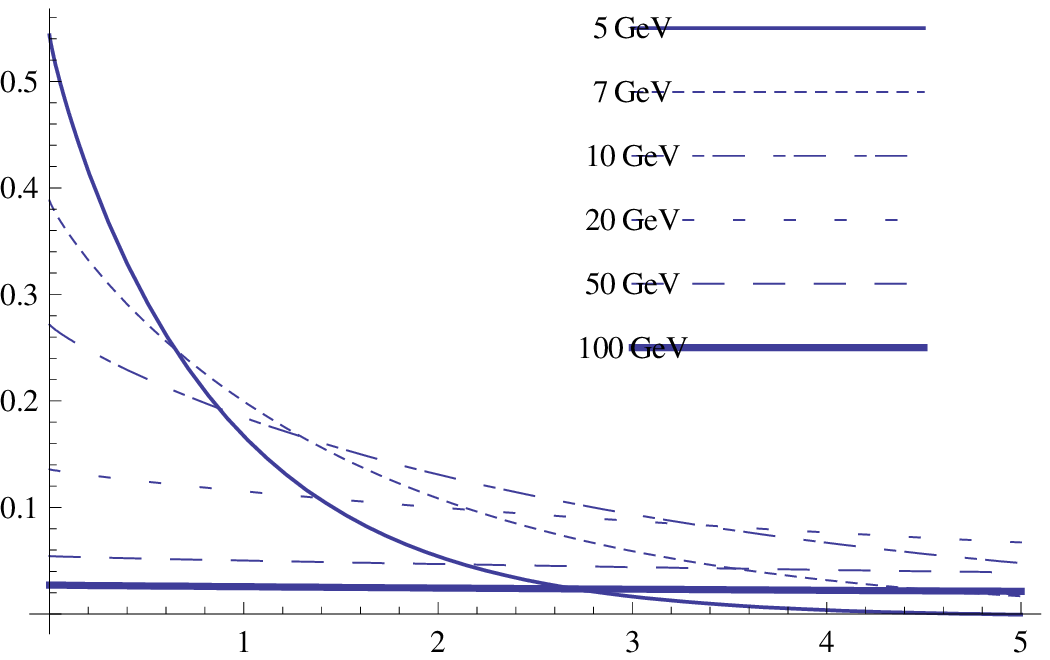}
}
\subfloat[]
{
\rotatebox{90}{\hspace{0.0cm} $d{\tilde H}/dQ\rightarrow$kg/(y keVee)}
\includegraphics[height=.17\textheight]{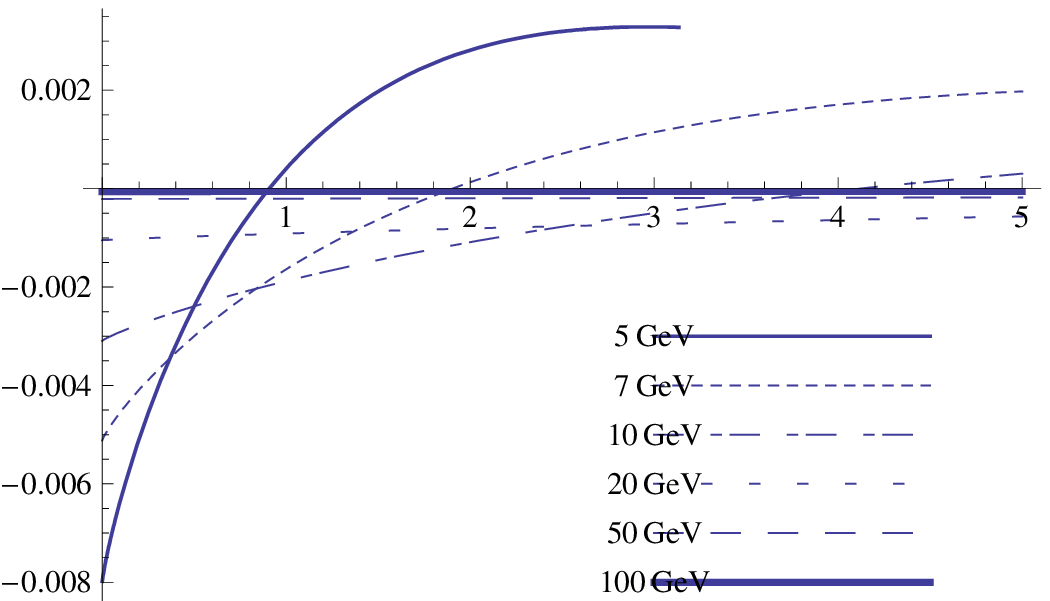}
}
\\
{\hspace{-2.0cm} $Q\rightarrow$keVee}
\caption{ The same as in Fig. \ref{fig:dRdQdHdQ_127} for the target $^{23}$Na.
 \label{fig:dRdQdHdQ_23}}
\end{center}
\end{figure}

We will apply the above formalism in the case of NaI, a target used in the DAMA experiment \cite{DAMA1,DAMA11}. The results for the Xe target are similar \cite{XENON10}. 
The differential rates $\frac{dR}{dQ}|_A$ and  $\frac{d\tilde{H}}{dQ}|_A$, for each component ($A=127$ and $A=23$) are exhibited in Fig. \ref{fig:dRdQdHdQ_127}-\ref{fig:dRdQdHdQ_23}. Following the practice of the DAMA experiment we express the energy transfer is in keVee using the phenomenological quenching factor \cite{LIDHART}, \cite{SIMON03} shown in Fig. \ref{fig:quenchfac}. The nuclear form factor has been included (for the $^{127}$I its effect is sizable even for an energy transfer of 10 keV, see Fig. \ref{fig:formfac}) .
The  differential rate for the spin mode for low energy transfers is similar to those exhibited in Figs \ref{fig:dRdQdHdQ_127}-\ref{fig:dRdQdHdQ_23}, since the spin form factors are similar. They are, of course, simply scaled down by $A^2$, if one takes the spin cross section, a combination of the nuclear spin ME and the nucleon spin amplitudes, to be the same with the coherent nucleon cross section, i.e.  $\sigma^{\mbox{{\tiny spin}}}_{\mbox{{\tiny nuclear}}}=10^{-7}$pb. For the actual spin nucleon cross sections extracted from experiment see \cite{JDVSpin09} and \cite{PICASSO09,COUPP11,SIMPLE11}.
 
\begin{figure}
\begin{center}
\subfloat[]
{
\rotatebox{90}{\hspace{0.0cm} $H(a \sqrt{u}) \cos{\alpha}\rightarrow$}
\includegraphics[height=.15\textheight]{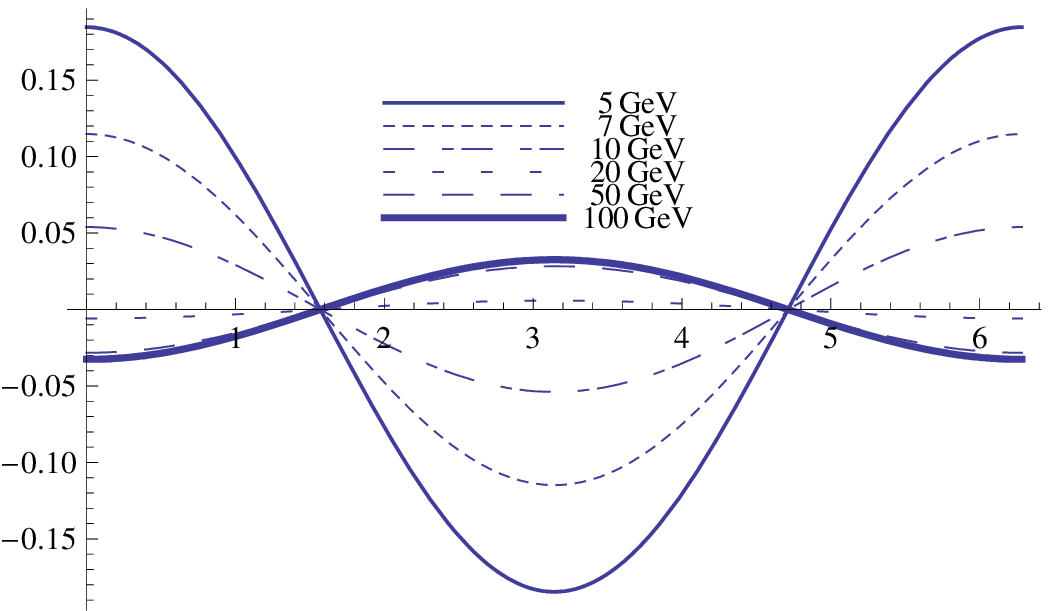}
}
\subfloat[]
{
\rotatebox{90}{\hspace{0.0cm} $H(a \sqrt{u}) \cos{\alpha}\rightarrow$}
\includegraphics[height=.15\textheight]{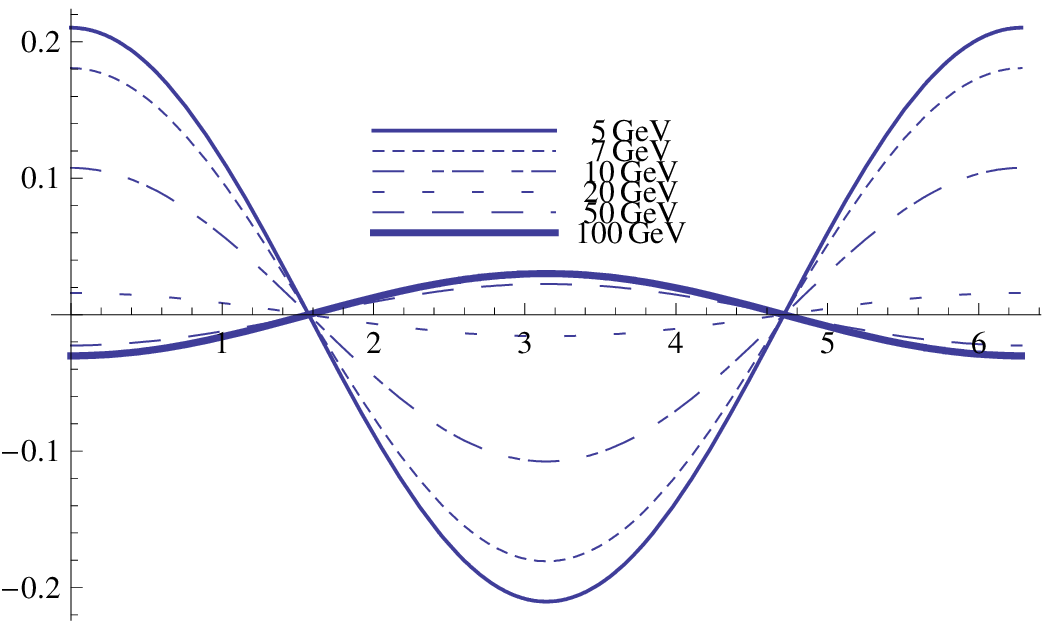}
}
\\
{\hspace{-2.0cm} $\alpha \rightarrow$}
\caption{ The modulation $H(a \sqrt{u}) \cos{\alpha}$ with an energy transfer of 1 keVee (a) and 2 keVee (b) for a heavy target (I or Xe). The solid, dotted, dot-dashed, dashed, long dashed and thick solid lines correspond to 5, 7, 10, 20, 50 and 100 GeV WIMP masses. Note that for some wimp masses on June 2nd the amplitude becomes negative (location of minimum rate). Note that the modulation is given relative to the time averaged rate.
 \label{fig:Hcosa127a}}
\end{center}
\end{figure}

\begin{figure}
\begin{center}
\subfloat[]
{
\rotatebox{90}{\hspace{0.0cm} $H(a \sqrt{u}) \cos{\alpha}\rightarrow$}
\includegraphics[height=.15\textheight]{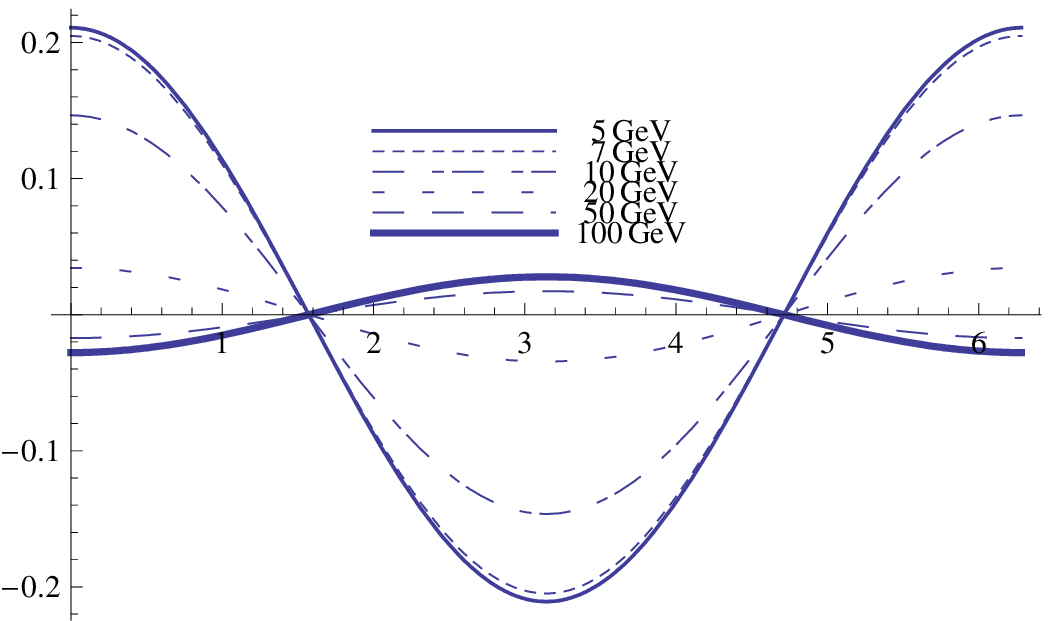}
}
\subfloat[]
{
\rotatebox{90}{\hspace{0.0cm} $H(a \sqrt{u}) \cos{\alpha}\rightarrow$}
\includegraphics[height=.15\textheight]{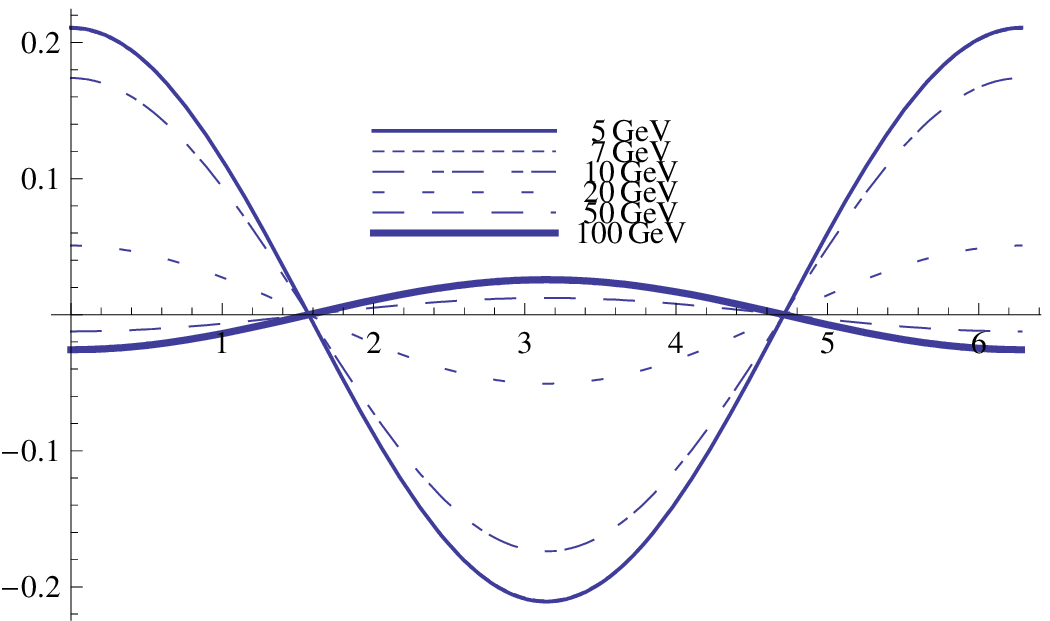}
}
\\
{\hspace{-2.0cm} $\alpha \rightarrow$}
\caption{ The same as in Fig. \ref{fig:Hcosa127a} with an energy transfer of 3 keVee (a) and 4 keVee (b).
 \label{fig:Hcosa127b}}
\end{center}
\end{figure}

\begin{figure}
\begin{center}
\subfloat[]
{
\rotatebox{90}{\hspace{0.0cm} $H(a \sqrt{u}) \cos{\alpha}\rightarrow$}
\includegraphics[height=.15\textheight]{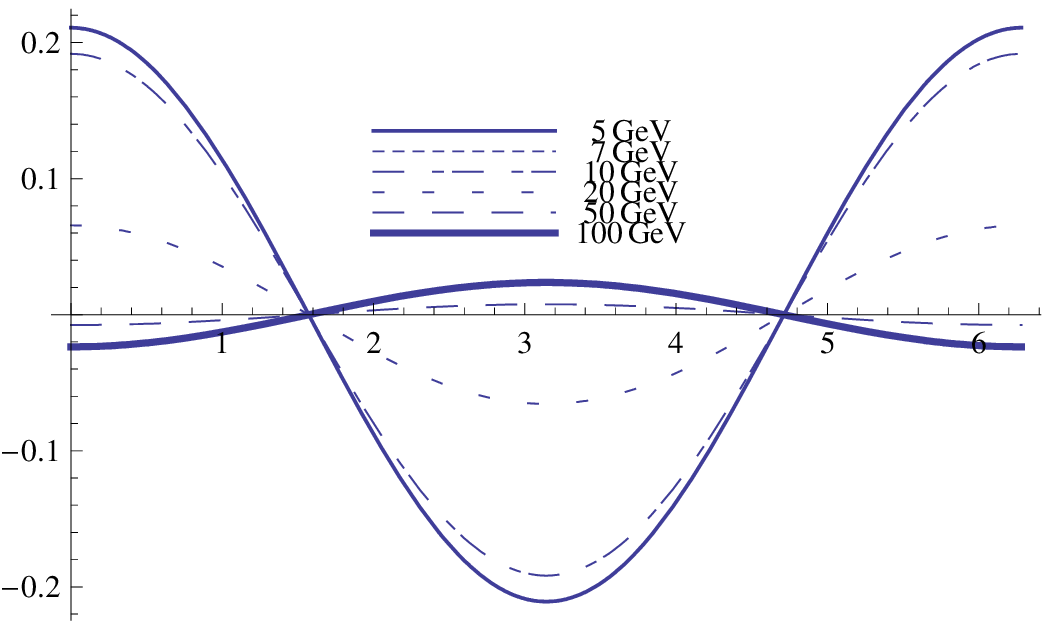}
}
\subfloat[]
{
\rotatebox{90}{\hspace{0.0cm} $H(a \sqrt{u}) \cos{\alpha}\rightarrow$}
\includegraphics[height=.15\textheight]{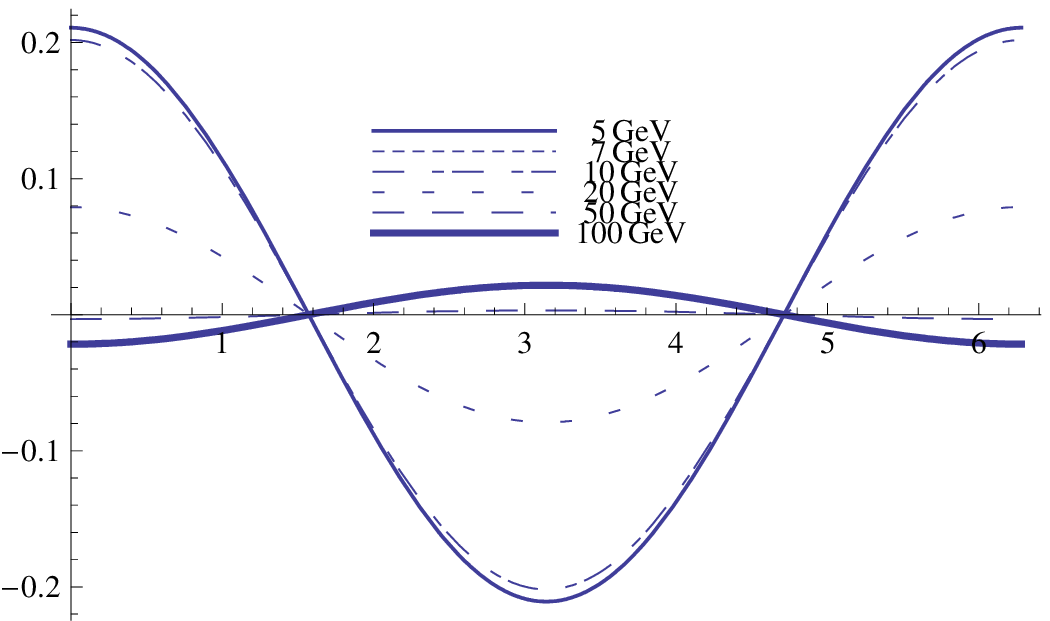}
}
\\
{\hspace{-2.0cm} $\alpha \rightarrow$}
\caption{ The same as in Fig. \ref{fig:Hcosa127a} with an energy transfer of 5 keVee (a) and 6 keVee (b).
 \label{fig:Hcosa127c}}
\end{center}
\end{figure}

\begin{figure}
\begin{center}
\subfloat[]
{
\rotatebox{90}{\hspace{0.0cm} $H(a \sqrt{u}) \cos{\alpha}\rightarrow$}
\includegraphics[height=.15\textheight]{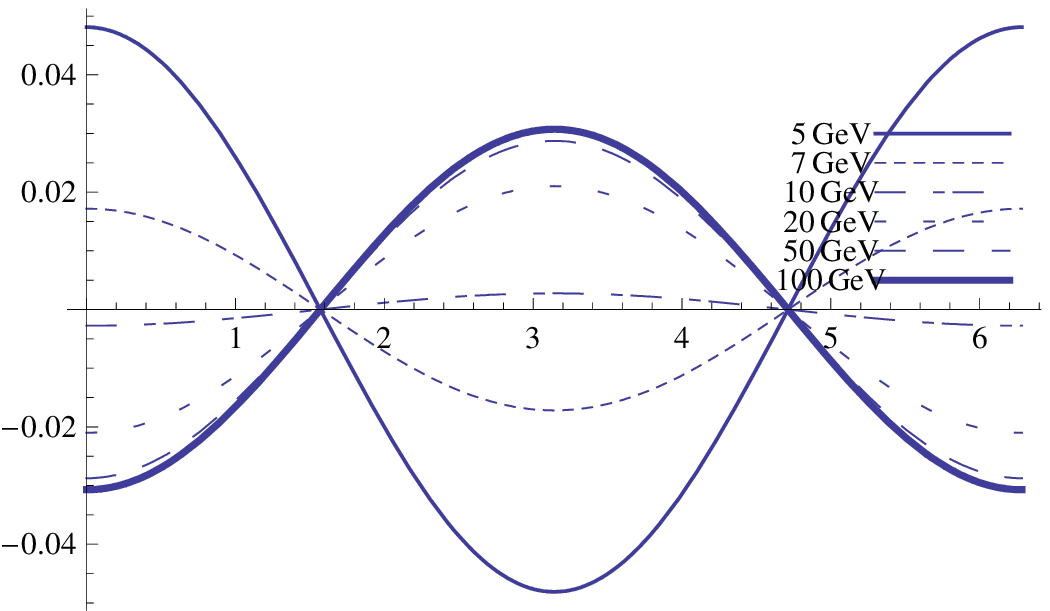}
}
\subfloat[]
{
\rotatebox{90}{\hspace{0.0cm} $H(a \sqrt{u}) \cos{\alpha}\rightarrow$}
\includegraphics[height=.15\textheight]{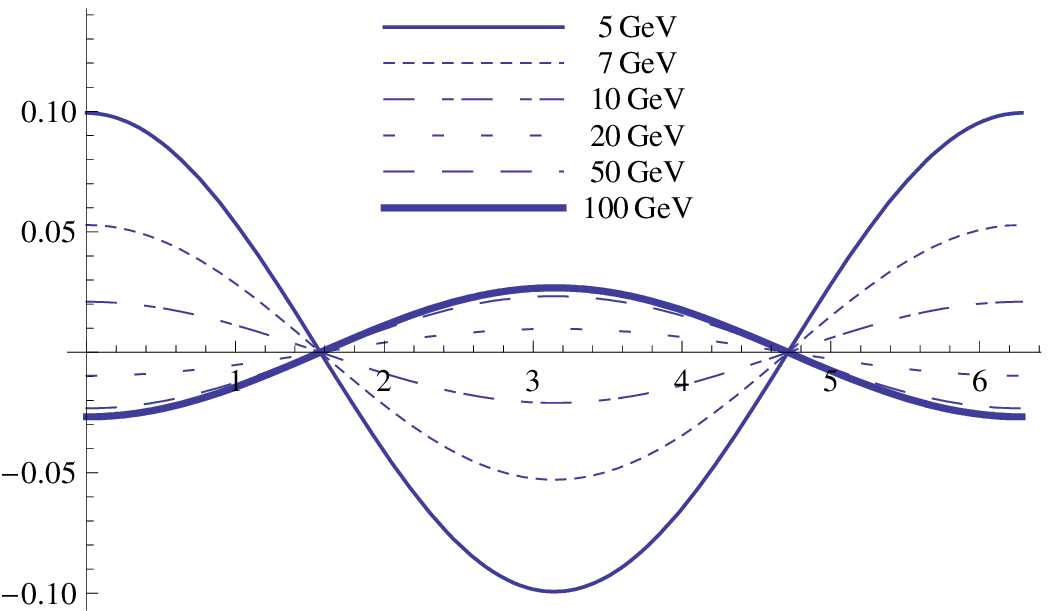}
}
\\
{\hspace{-2.0cm} $\alpha \rightarrow$}
\caption{ The same as in Fig. \ref{fig:Hcosa127a} for a light target (Na or F).
 \label{fig:Hcosa23a}}
\end{center}
\end{figure}

\begin{figure}
\begin{center}
\subfloat[]
{
\rotatebox{90}{\hspace{0.0cm} $H(a \sqrt{u}) \cos{\alpha}\rightarrow$}
\includegraphics[height=.15\textheight]{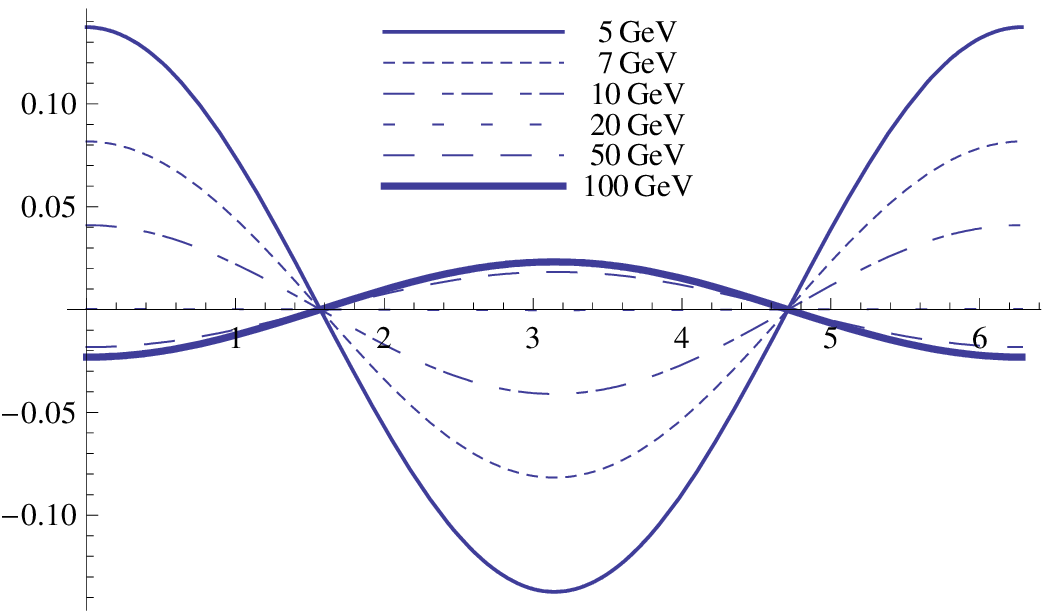}
}
\subfloat[]
{
\rotatebox{90}{\hspace{0.0cm} $H(a \sqrt{u}) \cos{\alpha}\rightarrow$}
\includegraphics[height=.15\textheight]{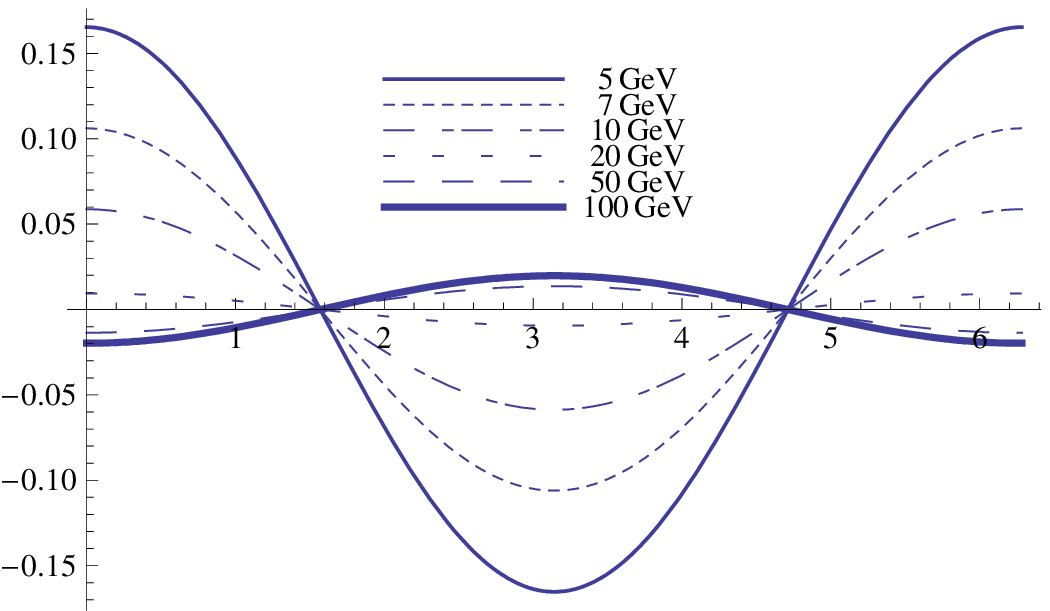}
}
\\
{\hspace{-2.0cm} $\alpha \rightarrow$}
\caption{ The same as in Fig. \ref{fig:Hcosa127b} with a light target (Na or F).
 \label{fig:Hcosa23b}}
\end{center}
\end{figure}

 The functions  $H(a \sqrt{u})\cos{\alpha}$  for each target component are shown in Figs  \ref{fig:Hcosa127a}- \ref{fig:Hcosa23b} as a function of $\alpha$ for various low energy transfers. The corresponding quantities for the spin mode are almost identical. We see that for certain values of the WIMP mass the modulation amplitude changes sign. This may perhaps by exploited to exreact information on the WIMP mass from the data. A similar behavior has been found by considering various halo models and different minimum WIMP velocities \cite{GREEN04,SFG06}.


\begin{figure}
\begin{center}
\subfloat[]
{
\rotatebox{90}{\hspace{0.0cm} $H(a \sqrt{u}) \cos{\alpha}\rightarrow$}
\includegraphics[height=.15\textheight]{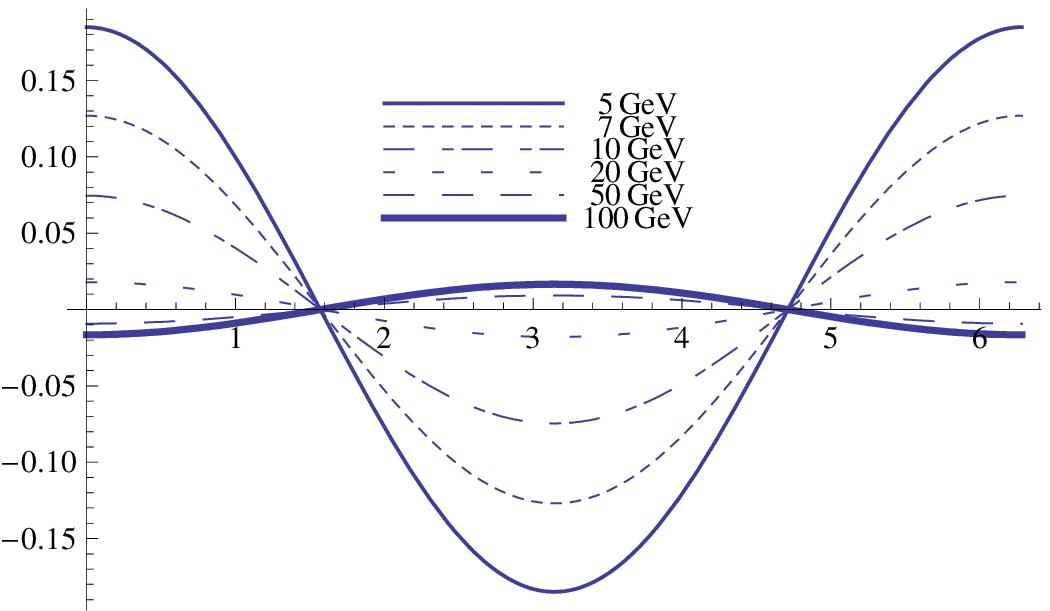}
}
\subfloat[]
{
\rotatebox{90}{\hspace{0.0cm} $H(a \sqrt{u}) \cos{\alpha}\rightarrow$}
\includegraphics[height=.15\textheight]{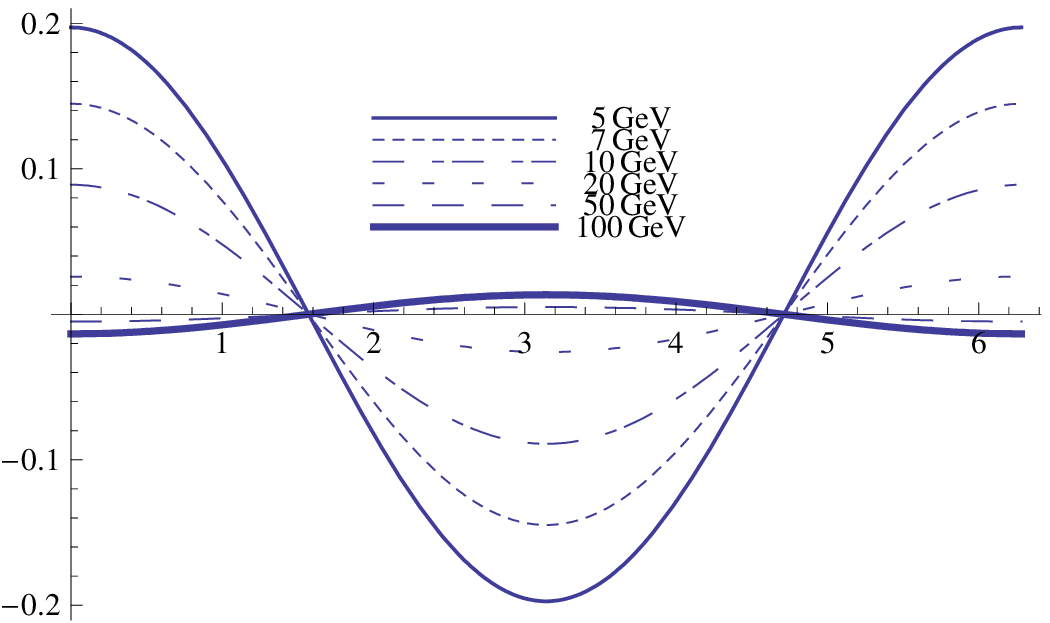}
}
\\
{\hspace{-2.0cm} $\alpha \rightarrow$}
\caption{ The same as in Fig. \ref{fig:Hcosa127c} with a light target (Na or F).
 \label{fig:Hcosa23c}}
\end{center}
\end{figure}
\begin{figure}
\begin{center}
\subfloat[]
{
\rotatebox{90}{\hspace{0.0cm} $dR/dQ\rightarrow$}
\includegraphics[height=.17\textheight]{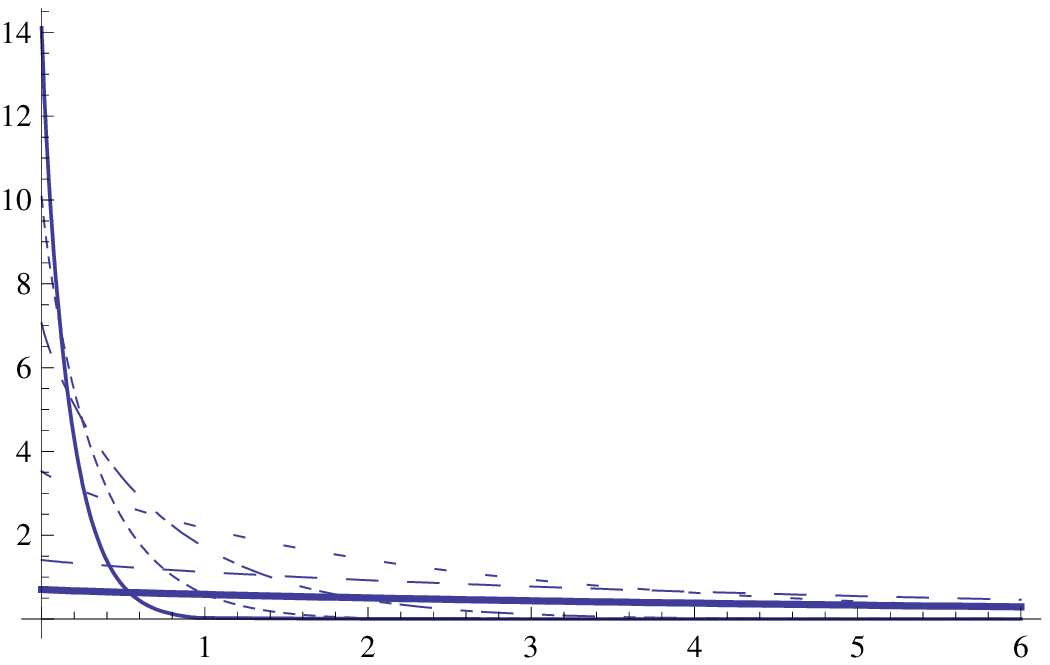}
}
\subfloat[]
{
\rotatebox{90}{\hspace{0.0cm} $d\tilde{H}/dQ\rightarrow$}
\includegraphics[height=.17\textheight]{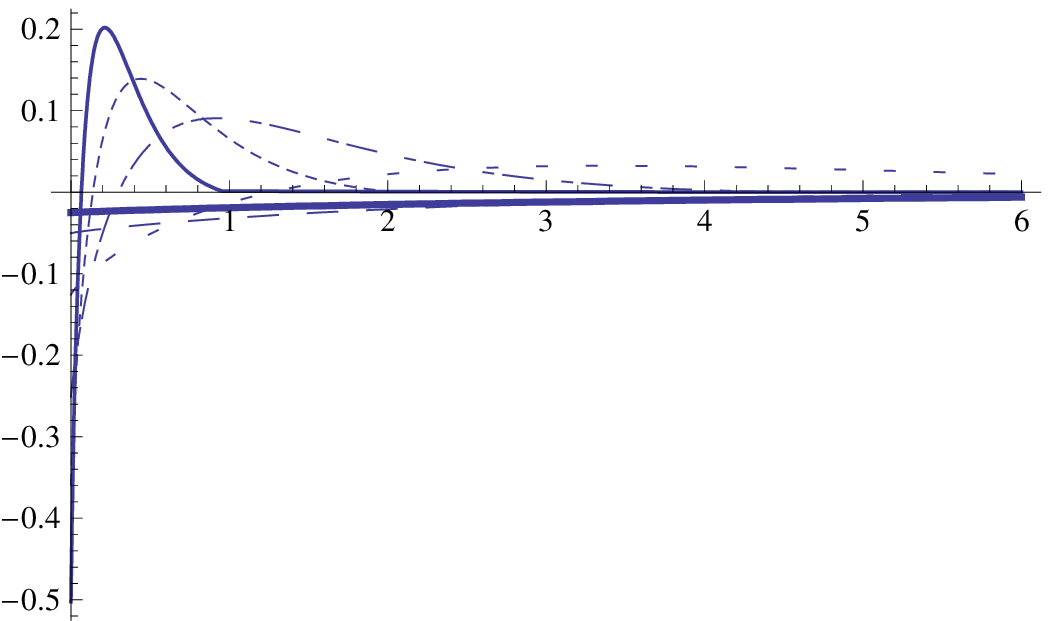}
}
\\
{\hspace{-2.0cm} $Q\rightarrow$keVee}
\caption{ The same as in Fig. \ref{fig:dRdQdHdQ_127} for the target NaI.
 \label{fig:dRdQdHdQ}}
\end{center}
\end{figure}
\begin{figure}
\begin{center}
\subfloat[]
{
\rotatebox{90}{\hspace{0.0cm} $H(a \sqrt{u}) \cos{\alpha}\rightarrow$}
\includegraphics[height=.15\textheight]{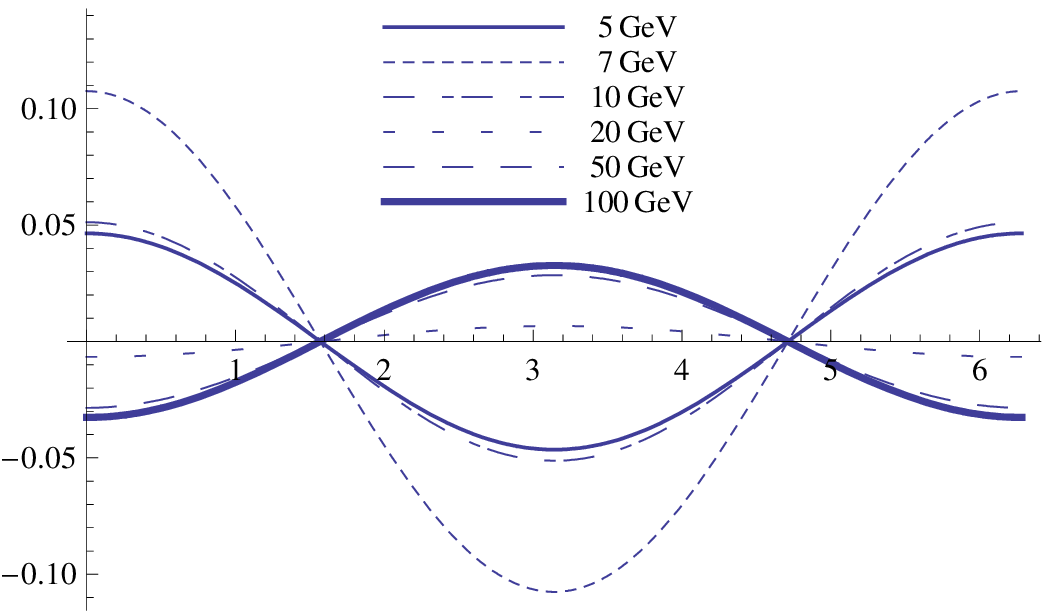}
}
\subfloat[]
{
\rotatebox{90}{\hspace{0.0cm} $H(a \sqrt{u}) \cos{\alpha}\rightarrow$}
\includegraphics[height=.15\textheight]{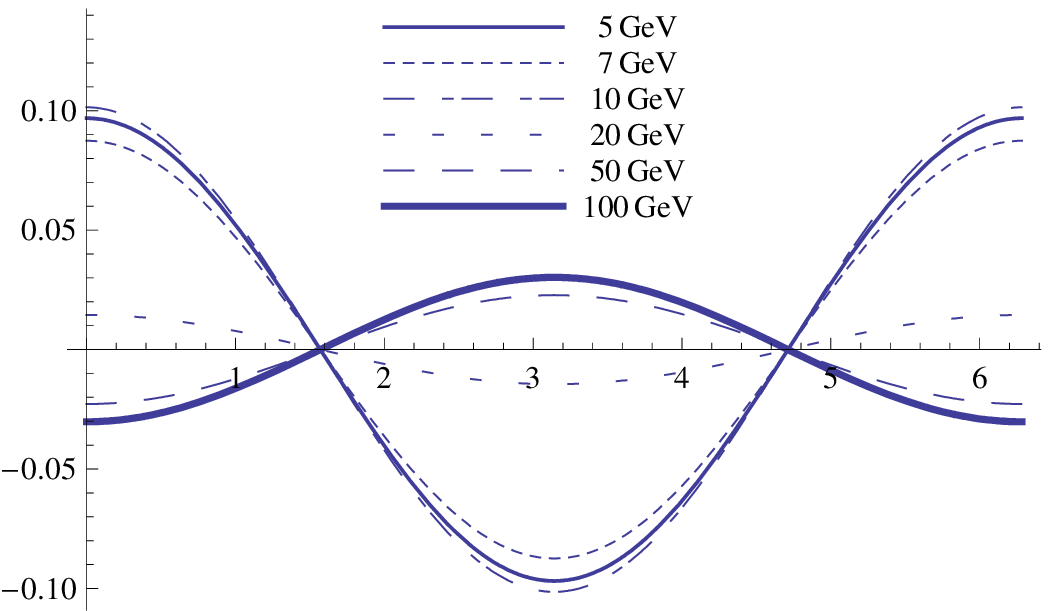}
}
\\
{\hspace{-2.0cm} $\alpha \rightarrow$}
\caption{ The same as in Fig. \ref{fig:Hcosa127a} for a NaI target.
 \label{fig:Hcosabotha}}
\end{center}
\end{figure}
\begin{figure}
\begin{center}
\subfloat[]
{
\rotatebox{90}{\hspace{0.0cm} $H(a \sqrt{u}) \cos{\alpha}\rightarrow$}
\includegraphics[height=.15\textheight]{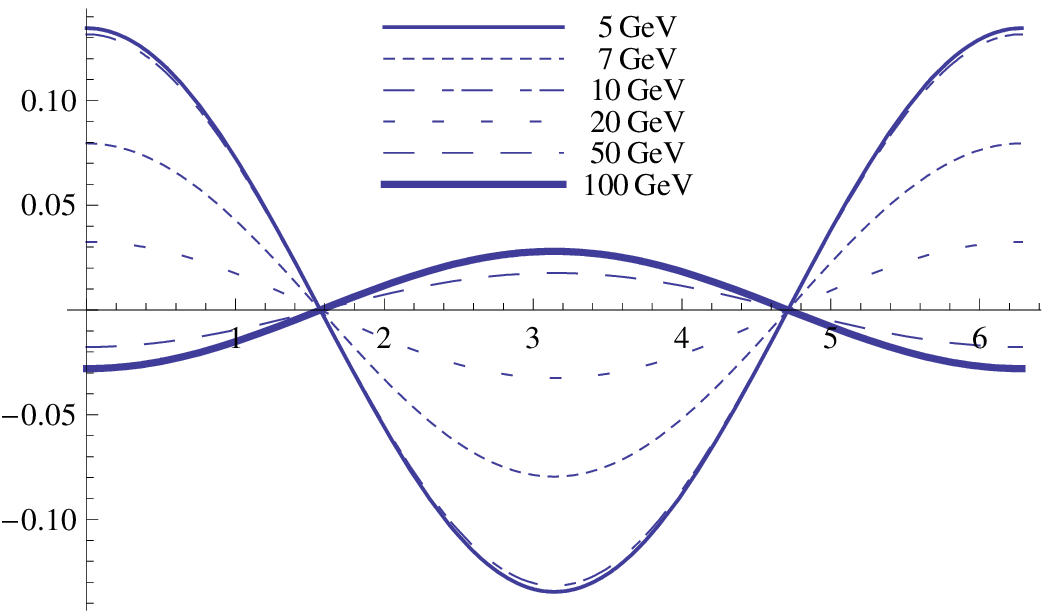}
}
\subfloat[]
{
\rotatebox{90}{\hspace{0.0cm} $H(a \sqrt{u}) \cos{\alpha}\rightarrow$}
\includegraphics[height=.15\textheight]{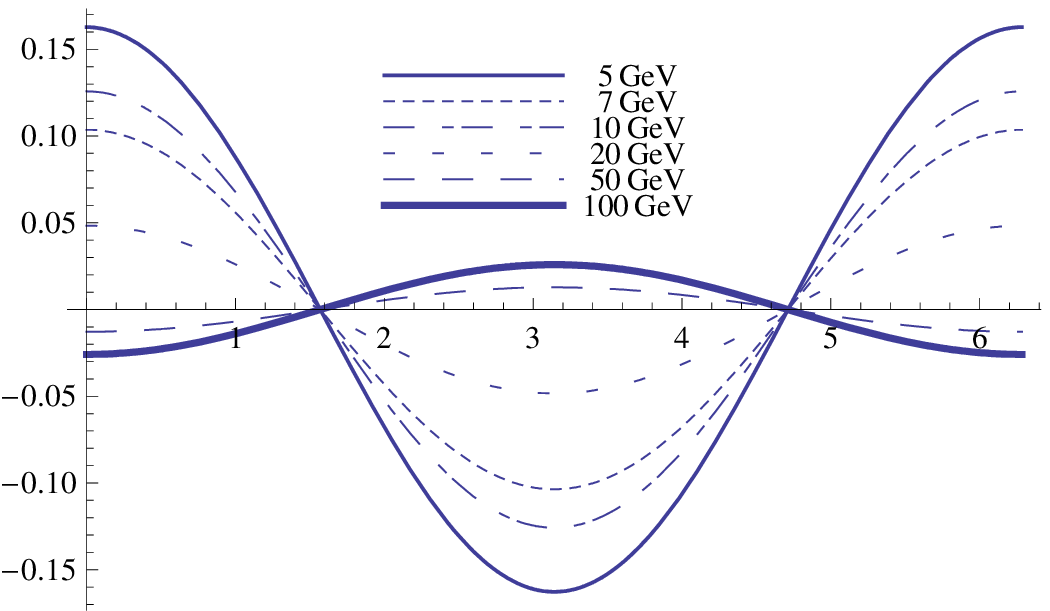}
}
\\
{\hspace{-2.0cm} $\alpha \rightarrow$}
\caption{ The same as in Fig. \ref{fig:Hcosa127b} with a  target of NaI.
 \label{fig:Hcosabothb}}
\end{center}
\end{figure}

Sometimes, as is the case for the DAMA experiment, the target has many components. In such cases the above formalism can be applied as follows:
\beq
\frac{dR}{dQ}|_A\rightarrow \sum_i X_i\frac{dR}{dQ}|_{A_i},\quad u\rightarrow u_i,\quad X_i=\mbox{the fraction of the component } A_i\mbox{ in the target}
\eeq
Thus we get the results shown in Figs \ref{fig:dRdQdHdQ} and \ref{fig:Hcosabotha}-\ref{fig:Hcosabothb}. The corresponding ones for the spin mode are not expected to be the same.


\begin{figure}
\begin{center}
\subfloat[]
{
\rotatebox{90}{\hspace{0.0cm} $H(a \sqrt{u}) \cos{\alpha}\rightarrow$}
\includegraphics[height=.15\textheight]{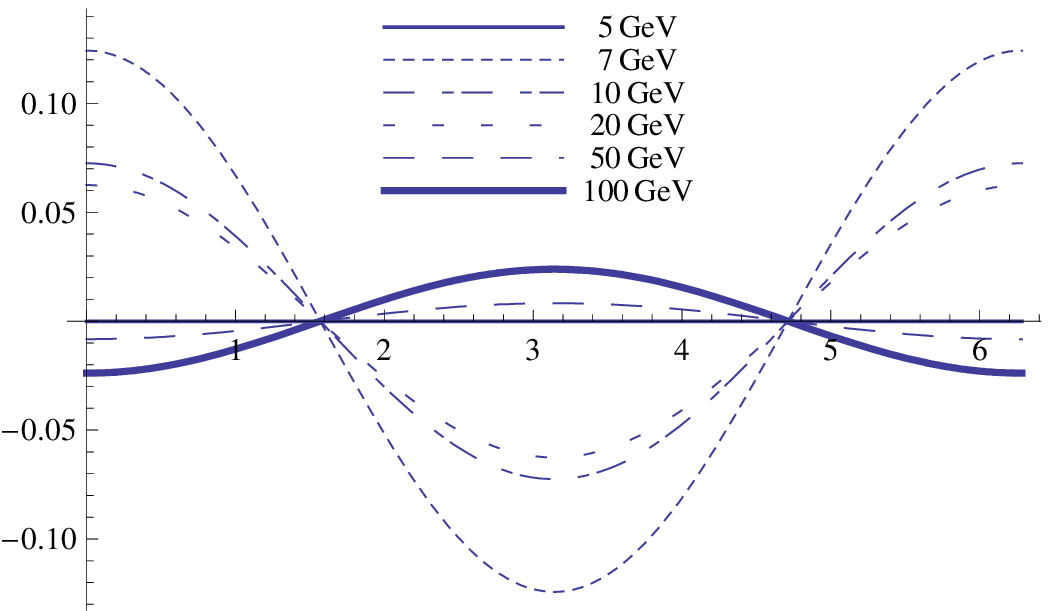}
}
\subfloat[]
{
\rotatebox{90}{\hspace{0.0cm} $H(a \sqrt{u}) \cos{\alpha}\rightarrow$}
\includegraphics[height=.15\textheight]{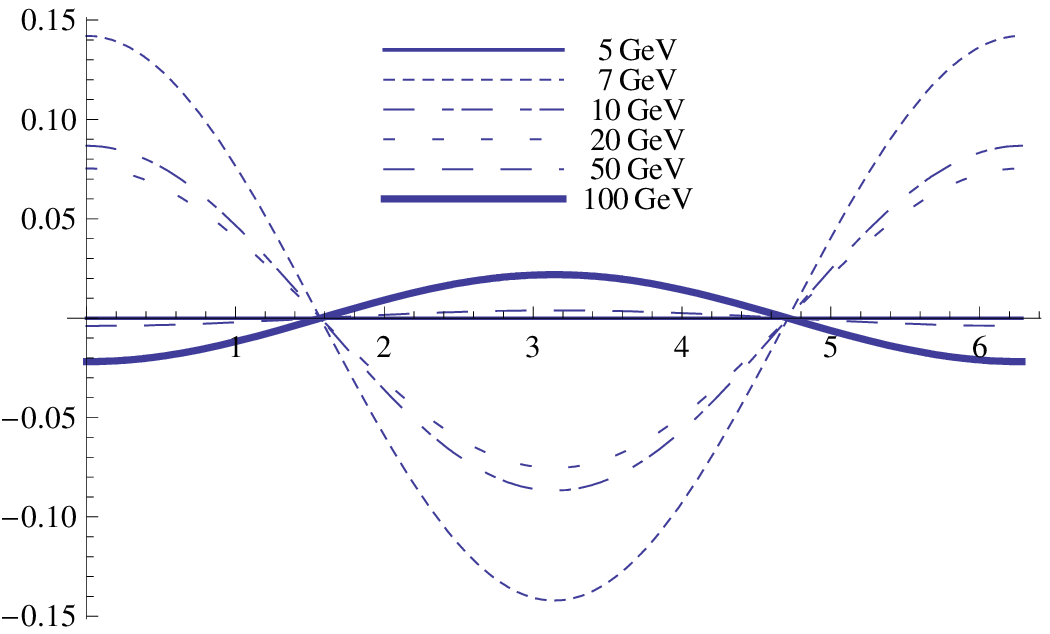}
}
\\
{\hspace{-2.0cm} $\alpha \rightarrow$}
\caption{ The same as in Fig. \ref{fig:Hcosa127c} with a  target of NaI.
 \label{fig:Hcosabothc}}
\end{center}
\end{figure}
\section{Some results on total rates}

\begin{figure}
\begin{center}
\subfloat[]
{
\rotatebox{90}{\hspace{0.0cm} $R\rightarrow$kg-y}
\includegraphics[height=.30\textwidth]{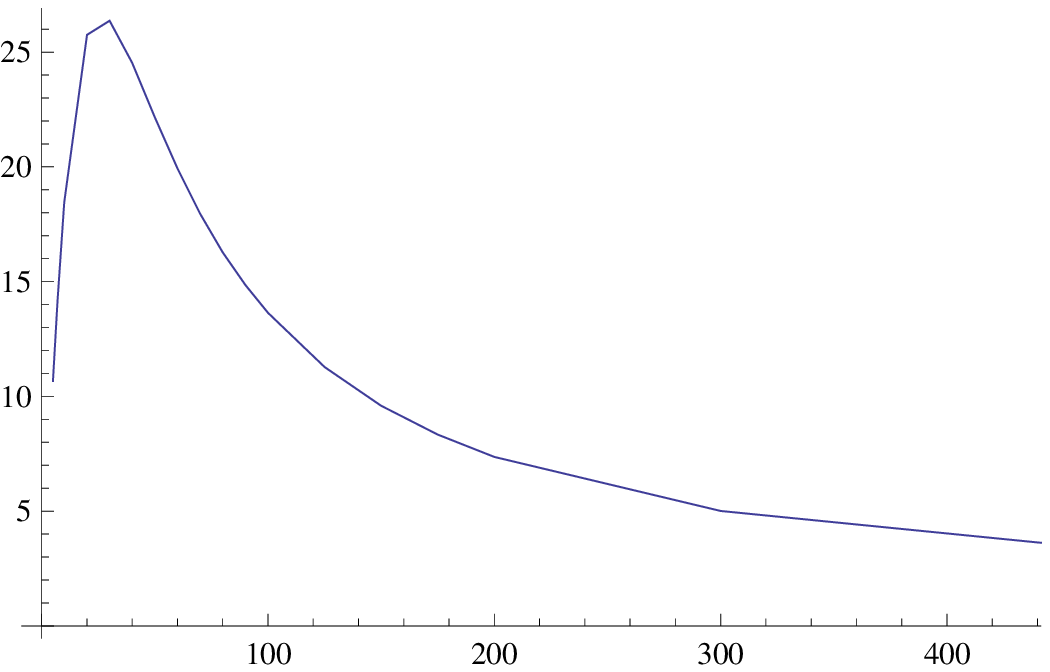}
}
\subfloat[]
{
\rotatebox{90}{\hspace{0.0cm} $R\rightarrow$kg}
\includegraphics[height=.30\textwidth]{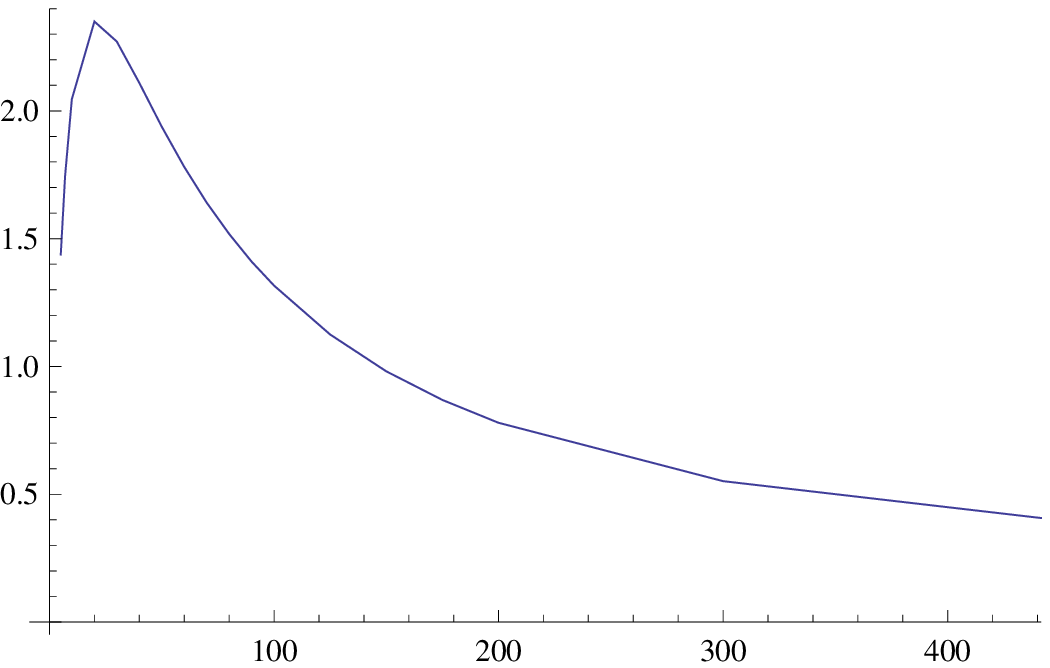}
}\\
{\hspace{-2.0cm} $m_{\text{WIMP}} \rightarrow$GeV}\\
\subfloat[]
{
\rotatebox{90}{\hspace{0.0cm} $R\rightarrow$kg}
\includegraphics[height=.50\textwidth]{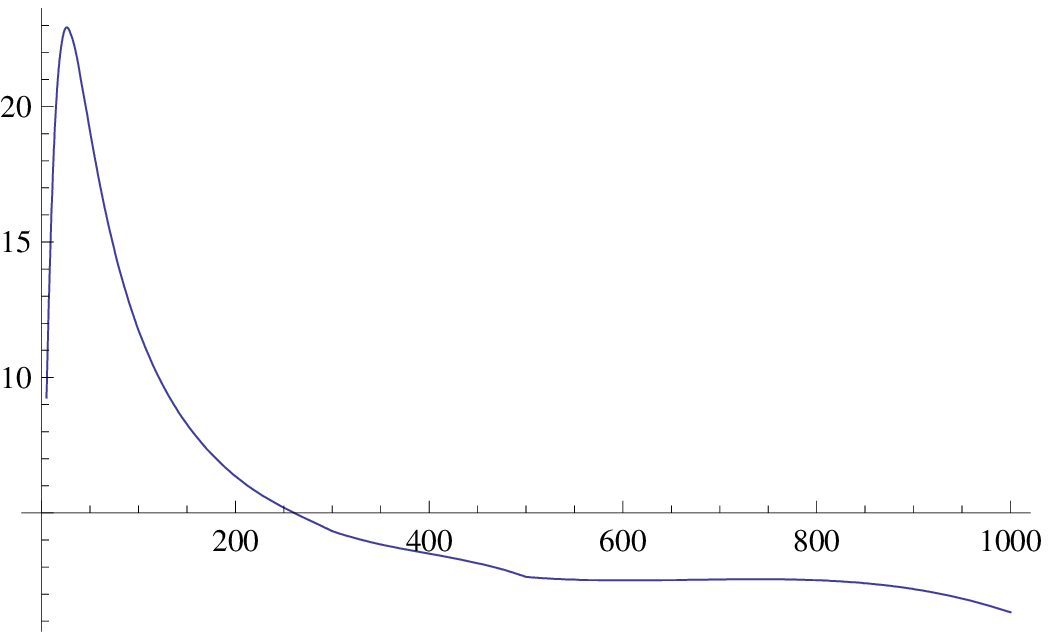}
}\\
{\hspace{-2.0cm} $m_{\text{WIMP}} \rightarrow$GeV}
\caption{ The total (time averaged) event rate in kg-y for a kg of target of $^{127}$I (a), of $^{23}$Na (b) and of NaI (c) assuming a coherent  nucleon cross section $\sigma_n=10^{-7}$pb and a zero threshold energy.
 \label{fig:Rtot}}
\end{center}
\end{figure}

\begin{figure}
\begin{center}
\subfloat[]
{
\rotatebox{90}{\hspace{0.0cm} $\tilde{H}\rightarrow$kg-y}
\includegraphics[height=.30\textwidth]{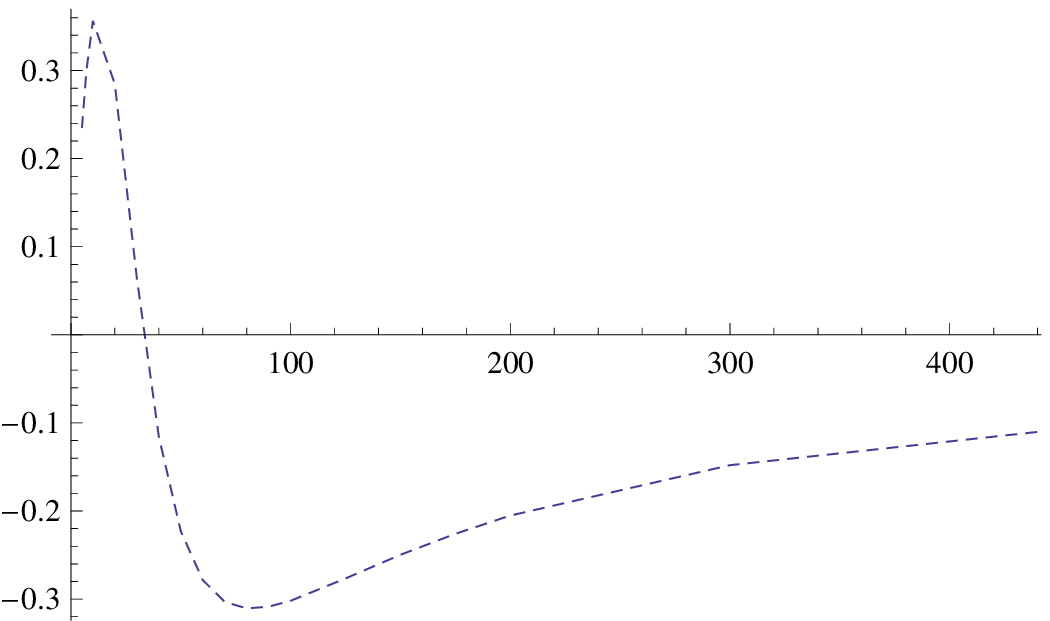}
}
\subfloat[]
{
\rotatebox{90}{\hspace{0.0cm} $\tilde{H}\rightarrow$kg-y}
\includegraphics[height=.30\textwidth]{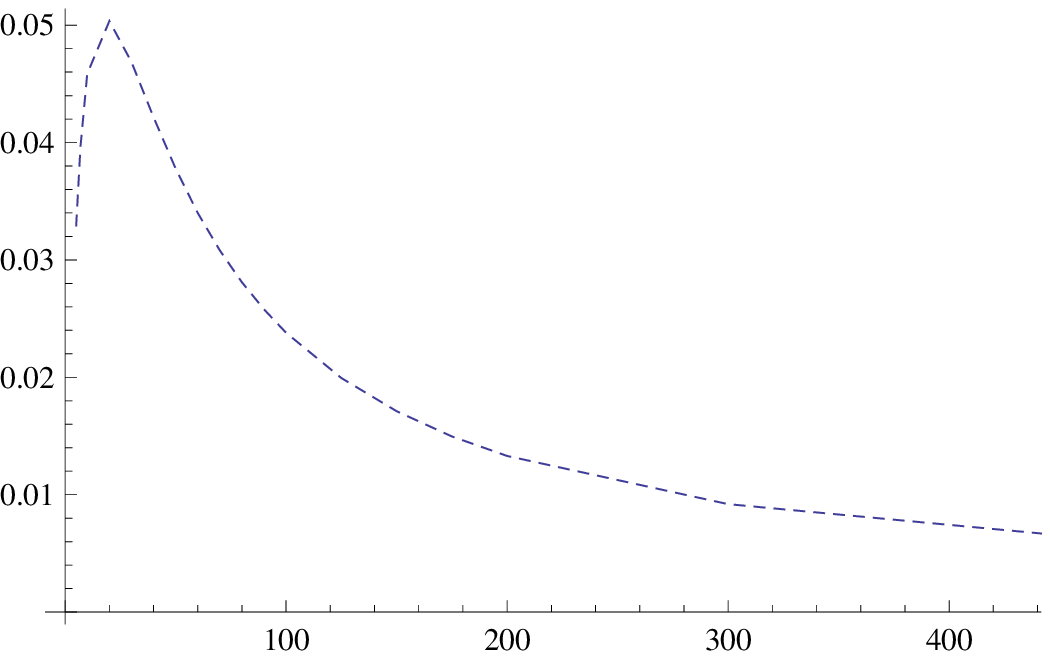}
}
\\
{\hspace{-2.0cm} $m_{\text{WIMP}} \rightarrow$GeV}\\
\subfloat[]
{
\rotatebox{90}{\hspace{0.0cm} $\tilde{H}\rightarrow$kg-y}
\includegraphics[height=.50\textwidth]{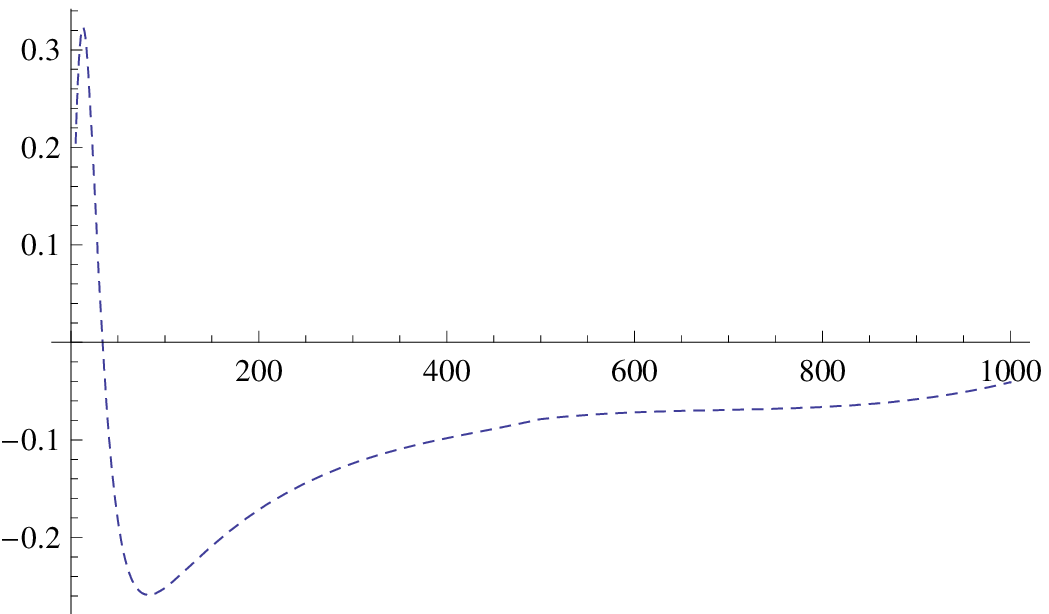}
}\\
{\hspace{-2.0cm} $m_{\text{WIMP}} \rightarrow$GeV}
\caption{ The total modulated event rate in kg-y for a kg of target of $^{127}$I (a), of $^{23}$Na (b) and of NaI (c) assuming a coherent  nucleon cross section $\sigma_n=10^{-7}$pb and a zero threshold energy.
 \label{fig:Htot}}
\end{center}

\end{figure}

For completeness and comparison we will briefly present our results on the total rates. Integrating the differential rates discussed in the previous section we obtain the total time averaged rate $R_0$, the total modulated rate $\tilde{H}$ and the relative modulation amplitude$h$  given by:
\beq
R=R_0+\tilde{H} \cos{\alpha},\quad 
R=R_0\left (1+h \cos{\alpha}\right )
\eeq
 These are exhibited for zero threshold as functions of the WIMP mass in Figs \ref{fig:Rtot} and \ref{fig:Htot} respectively. Some special results in the case of low WIMP mass are exhibited in Tables \ref{tab1}-\ref{tab2}. In the case of non zero threshold one notices the strong dependence of the rime averaged rate on the WIMP mass. Also in this case the relative modulation $h$ substantially increases, the difference between the maximum and the minimum can reach 20$\%$. This however occurs at the expense of the number of counts, since both the time averaged and the time dependent part decrease, but the time averaged part decreases faster. So their ratio increases. This can be understood by noticing that the cancellation of the negative and positive parts in the differential modulated amplitide, see Fig. \ref{fig:psix}, becomes less effective in this case.

\begin{figure}
\begin{center}
\subfloat[]
{
\rotatebox{90}{\hspace{0.0cm} $R\rightarrow$kg-y}
\includegraphics[height=.30\textwidth]{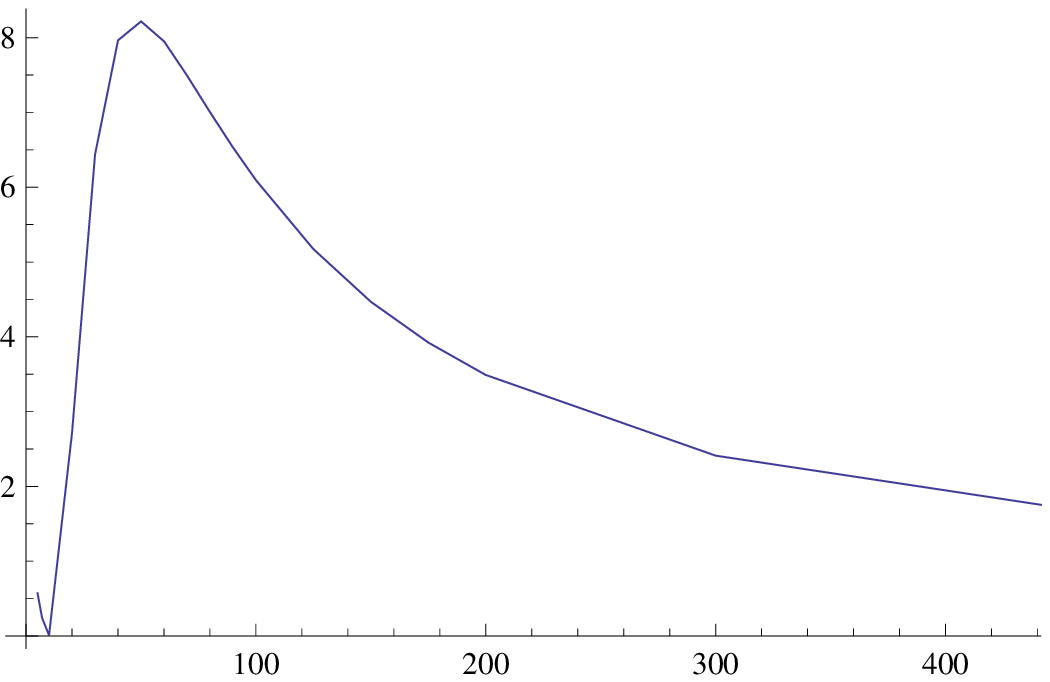}
}
\subfloat[]
{
\rotatebox{90}{\hspace{0.0cm} $R\rightarrow$kg}
\includegraphics[height=.30\textwidth]{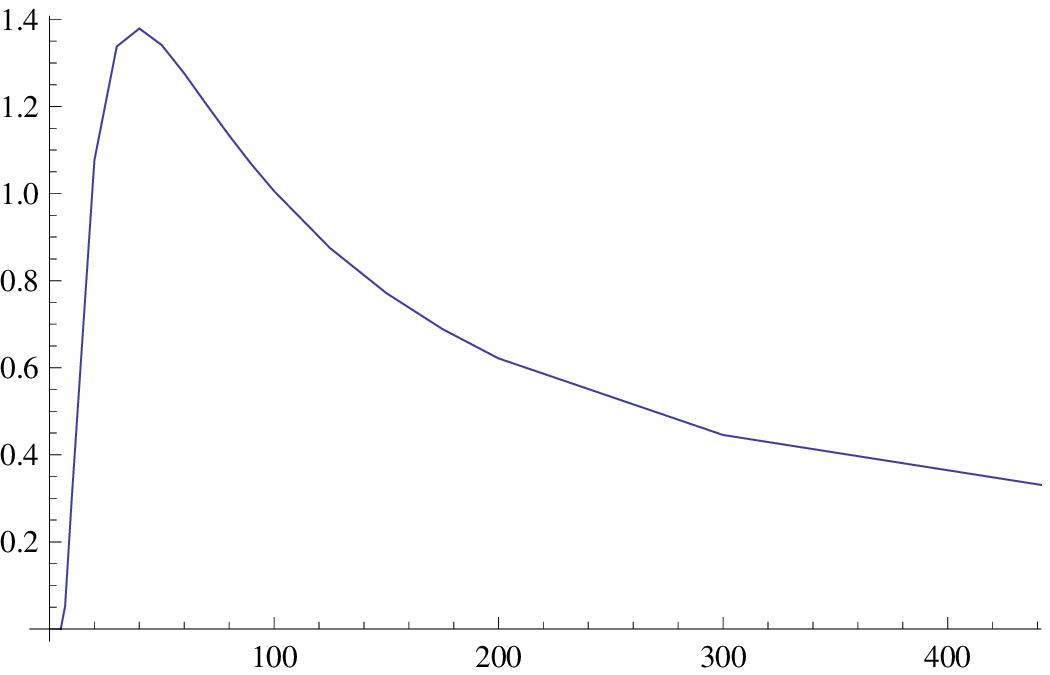}
}\\
{\hspace{-2.0cm} $m_{\text{WIMP}} \rightarrow$GeV}\\
\subfloat[]
{
\rotatebox{90}{\hspace{0.0cm} $R\rightarrow$kg}
\includegraphics[height=.50\textwidth]{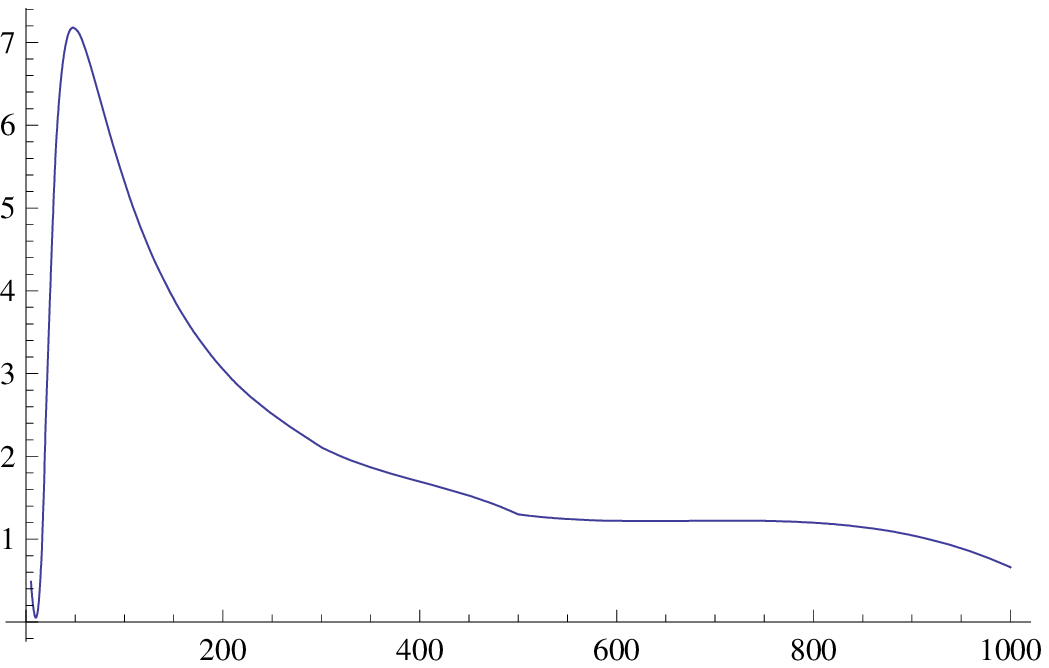}
}\\
{\hspace{-2.0cm} $m_{\text{WIMP}} \rightarrow$GeV}
\caption{ The total (time averaged) event rate in kg-y for a kg of target of $^{127}$I (a), of $^{23}$Na (b) and of NaI (c) assuming a coherent  nucleon cross section $\sigma_n=10^{-7}$pb and a  threshold energy of 5 keVee.
 \label{fig:Rtot_5}}
\end{center}
\end{figure}
\begin{table}
\begin{center}
  \caption{Some the total event rates for some special WIMP masses and energy thresholds. The coherent nucleon cross section of $\sigma_n=10^{-7}$pb was employed.}
  \label{tab1}
  \begin{tabular}{|l|l|l|l|l|l|l|l|l|l|l|}
  \hline\hline
  & & & & & & & & & &\\
   $E_{th}$&$m_{\text{WIMP}}$  & $R_0$(I) & $\tilde{H}$(I) &h(I)& $R_0$(Na) & $\tilde{H}$(Na) &h(Na)& $R_0$(NaI) & $\tilde{H}$(NaI) &h(NaI)\\
       (keVee)& GeV &kg-y  &kg-y  & & kg-y & kg-y & &kg-y  & kg-y &\\
   \hline
 0 & 80 & 16.3 & -0.311 & -0.019 & 1.518 &
   0.028& 0.019 & 14.0 & -0.259& -0.018 \\
 0 & 20 & 25.8 & 0.285 & 0.011 & 2.35 & 0.050
   & 0.021& 22.2 & 0.249 & 0.019 \\
 0 & 10 & 18.4 & 0.356 & 0.019 & 2.045 & 0.046
   & 0.022& 15.9 & 0.309 & 0.019 \\
   \hline
 5 & 80 & 7.00 & -0.042 & -0.006 & 1.133 &
   0.038 & 0.034 & 6.11 & -0.030 & -0.005
   \\
 5 & 20 & 2.72 & 0.247 & 0.091 & 1.07 & 0.065
   & 0.060 & 2.47 & 0.219& 0.089 \\
 5 & 10 & 0.008 & 0.001 & 0.187 & 0.303 &
   0.031 & 0.103 & 0.053 & 0.006 & 0.114\\
   \hline
   \hline
   \end{tabular}
\end{center}
\end{table}
\begin{table}
\begin{center}
  \caption{The same as in table \ref{tab1} for $\sigma_n=2 \times 10^{-4}$pb relevant for the DAMA region.}
  \label{tab2}
  \begin{tabular}{|l|l|l|l|l|l|l|l|l|l|l|}
  \hline\hline
  & & & & & & & & & &\\
   $E_{th}$&$m_{\text{WIMP}}$  & $R_0$(I) & $\tilde{H}$(I) &h(I)& $R_0$(Na) & $\tilde{H}$(Na) &h(Na)& $R_0$(NaI) & $\tilde{H}$(NaI) &h(NaI)\\
       (keVee)& GeV &kg-y  &kg-y  & & kg-y & kg-y & &kg-y  & kg-y &\\
   \hline
 0 & 80 & $4.07\times 10^{4}$ & -776 & -0.019& $3.80\times 10^{3}$  & 70.2 &
   0.019 & $3.50\times 10^{4}$  & -647 & -0.018 \\
 0 & 20 & $6.43\times 10^{4}$  & 712 & 0.011 &$5.87\times 10^{3}$& 126 &
   0.021 & $5.54\times 10^{4}$ & 622 & 0.011 \\
 0 & 10 &$4.61\times 10^{4}$  & 891& 0.019 & $5.11\times 10^{3}$  & 115 &
   0.022 &$3.98\times 10^{4}$  & 772 & 0.019 \\
 5 & 80 & $1.75\times 10^{4}$  & -105 & -0.006 & $4.83\times 10^{3}$  & 95.0 &
   0.034 & $1.53\times 10^{4}$  & -74.6 & -0.005 \\
 5 & 20 & $6.80\times 10^{3}$ & 617 & 0.091& $2.69\times 10^{3}$  & 162 &
   0.060 & $6.17\times 10^{3}$  & 547 & 0.089 \\
 5 & 10 & 19.4 & 3.62 & 0.187 & 757 & 78.1 &
   0.103& 132 & 15.0 & 0.114\\
   \hline
   \hline
   \end{tabular}
\end{center}
\end{table}

\begin{figure}
\begin{center}
\subfloat[]
{
\rotatebox{90}{\hspace{0.0cm} $\tilde{H}\rightarrow$kg-y}
\includegraphics[height=.30\textwidth]{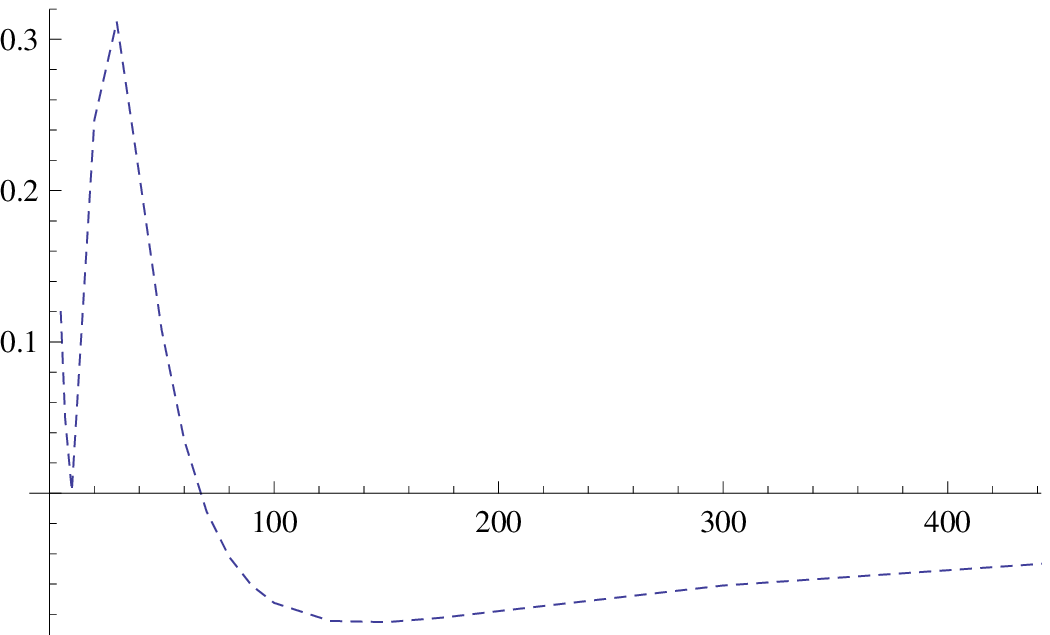}
}
\subfloat[]
{
\rotatebox{90}{\hspace{0.0cm} $\tilde{H}\rightarrow$kg-y}
\includegraphics[height=.30\textwidth]{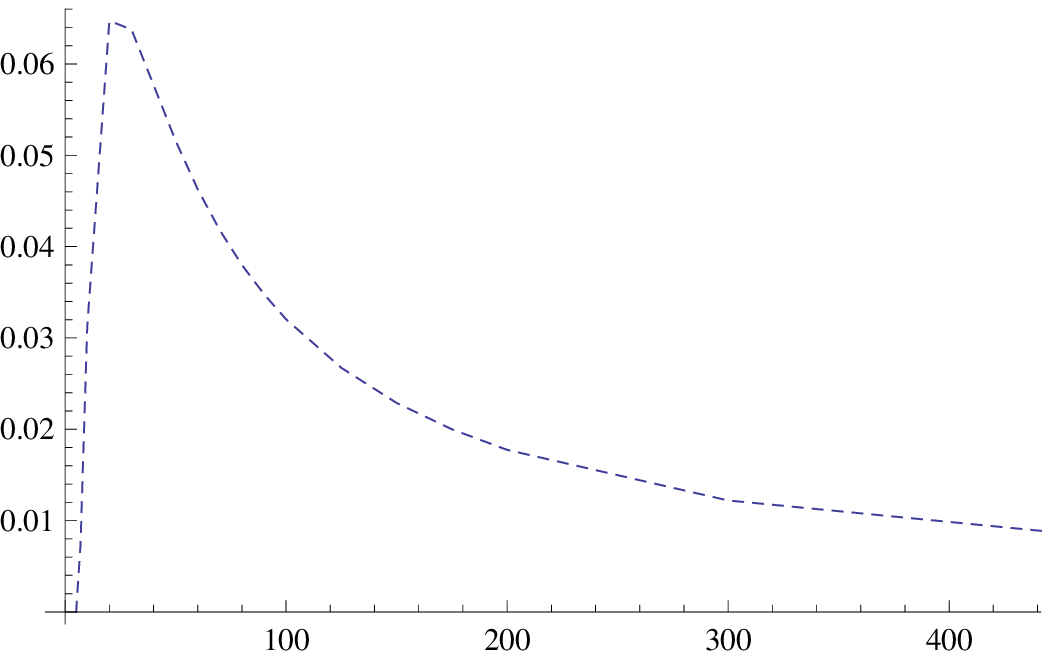}
}
\\
{\hspace{-2.0cm} $m_{\text{WIMP}} \rightarrow$GeV}\\
\subfloat[]
{
\rotatebox{90}{\hspace{0.0cm} $\tilde{H}\rightarrow$kg-y}
\includegraphics[height=.50\textwidth]{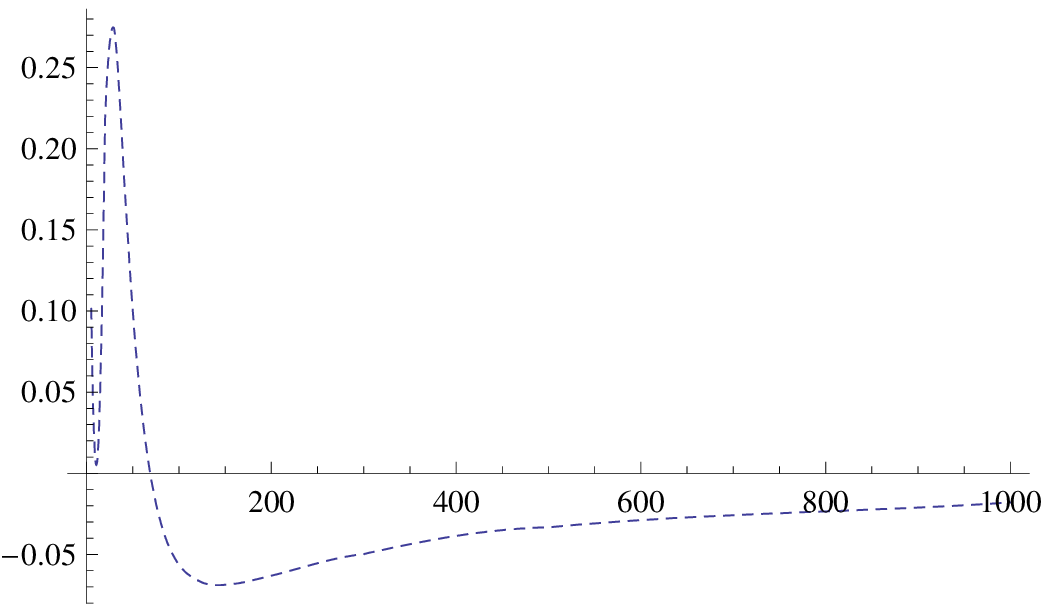}
}\\
{\hspace{-2.0cm} $m_{\text{WIMP}} \rightarrow$GeV}
\caption{ The total modulated event rate in kg-y for a kg of target of $^{127}$I (a), of $^{23}$Na (b) and of NaI (c) assuming a coherent  nucleon cross section $\sigma_n=10^{-7}$pb and a  threshold energy of 5 keVee.
 \label{fig:Htot_5}}
\end{center}
\end{figure}
\section{Discussion}
In the present paper we obtained results on the differential event rates, both modulated and time averaged, focusing our attention on small energy transfers  and relatively light WIMPS. We found that:
\begin{itemize}
\item The relative modulation amplitude crucially depends on the WIMP mass. For small masses it exhibits normal behavior, but for large masses it changes sign (minimum in June). This effect is more pronounced in the case of heavy targets.
\item The relative modulation amplitude depends somewhat on the energy transfer, especially at low transfers.
\item For WIMP masses less than 10 GeV, the difference between the maximum and the minimum could reach between $20\%$ and $40\%$ for a heavy target, but it is a bit less for a light target, depending on the enegy transfer.
\item The relative modulation amplitude for NaI is the weighted average of its two components, and in the low energy regime, between 1 and 6 keVee, it does not change much with the energy transfer.
\item Once it is established that one actually observes the modulation effect, the sign of the modulation may be exploited to infer the WIMP mass.
\end{itemize}
For low WIMP mass  the total rates depend strongly on the threshold energy, especially for a heavy target. The relative modulation in the presence a threshold gets quite large ($h\approx 0.2$), but, unfortunately, this occurs at the expense of the number of counts. It is important to compare the relative total modulation in a least one light and one heavy target. For very low energy thresholds, if the signs are opposite, one may infer that the WIMP is heavy, $m_{\mbox{\tiny{WIMP}}}\ge100$ GeV.

\end{document}